\documentclass[preprint,aps,prb,showpacs,floatfix]{revtex4-2}

\usepackage{dcolumn}
\usepackage{bm}%
\usepackage{graphicx}
\usepackage{subfig}
\usepackage{float}

\def\dfrac#1#2{{\displaystyle{#1\over#2}}}

\def\dfrac{\displaystyle\frac}

 \usepackage{color}

\usepackage{ulem} 
\usepackage{soul}

\begin{document}
\title{Effect of the surface  on the energy levels in PbS and PbSe  nanocrystals}
\author{O. Lundgren}
\affiliation{Woodside High School,199 Churchill Ave, Woodside, CA 94062, USA}
\author{A.V. Rodina}
\affiliation{A. F. Ioffe Physico-Technical
Institute, 194021, St.-Petersburg, Russia}
\author{Al. L. Efros}
\email{Corresponding author: alex.l.efros.civ@us.navy.mil}
\affiliation{U. S. Naval Research
Laboratory, Washington, DC 20375, USA}
\date{\today, \jobname , 
}

\begin{abstract}
The structure of the electron and hole quantum confinement  levels in spherical PbS and PbSe nanocrystals (NCs) is studied within the framework of a four-band effective-mass model. This level structure was initially calculated within the Kang--Wise model, which employs the standard boundary condition of vanishing carrier wave functions at the NC surface. This model, however, does not allow one to account for the effect of the NC surface on the level structure. Here, the effect of the NC surface is modeled using general boundary conditions (GBCs) imposed on the envelope wave functions at the surface. We show that the surface properties modify the size dependence of the electron and hole confinement levels and lead to an additional spin--orbit splitting of the $1P_{h,e}$ levels. Using the GBCs, we are able to quantitatively describe the size dependence of the lowest optical transitions in bare PbS and PbSe NCs and explain the differences between the absorption spectra of bare and Cl-passivated PbS NCs. \end{abstract}
\maketitle

\section{Introduction}

 The electronic structure of spherical PbS and PbSe nanocrystals (NCs) was first calculated by  Kang\&Wise within a four-band envelope function formalism.\cite{kang_wise} The four band effective mass model was shown to describe reasonably well the size dependence of the first exciton transitions. The calculations showed that for the PbS and PbSe NCs the spherical approximation with the averaged material parameters can be used and the effect of the anisotropy can be taken into account as a perturbation.\cite{kang_wise} However, the model fails to predict the energy of the second exciton transition in PbSe, unless one assumes the $1S_{e,h}$-$1P_{h,e}$ inter-band transitions to be allowed.\cite{brian} These transitions are  forbidden in the spherical approximation by the parity selection rules. In contrast to the statement of Ref. \cite{andreev99}, the transitions between the hole ground (first excited) state and hole first excited (ground) state remain forbidden even if one takes into account the mixing of the states with different energies by the anisotropic perturbation \cite{kang_wise} or considers the anisotropic four--band envelope function model.\cite{tudury} The effects of anisotropy on the energy spectra were found to be significant for the PbSe and especially for the PbTe NCs.\cite{tudury} The effects of anisotropy split the first excited energy level of the holes and thus decrease the calculated energy of the second exciton transition in comparison with the predictions  of the spherical approximation.\cite{tudury}  However, this anisotropy--induced splitting can not fully  account for the observed shift of the second exciton optical transition in PbSe NCs.  In PbS  NCs absorption spectra
measured at 4 K quantitatively agree with the atomistic calculations \cite{Poddubny2012} that take the valley anisotropy into account. This allows one to unambiguously ascribe the much-debated second absorption peak to the $1P_e-1P_h$ optical transitions and directly measure the splitting of excited $P$- states.\cite{Poddubny2017}

In the calculation of Refs. \cite{kang_wise,tudury} the standard  boundary conditions   assuming the vanishing of the envelope function at the NCs surface were used. The standard boundary conditions (SBCs), however, were shown by Ando and Mori \cite{Ando1982} to be a special case of the more general boundary conditions (GBCs) which envelope wave functions must satisfy.  These GBCs originate from the requirement of continuity of the current density normal to the surface or interface and include an energy and state independent transfer matrix that is a unique characteristic of the surface or interface.
In the case of infinitely high potential barriers, the GBCs are expressed as the vanishing of the current density normal to the surface (see for example Refs. \onlinecite{brag99,bc}).  As a result, the envelope wave function  does not  vanish  at the surface and satisfies the GBCs that are the characteristics of the particular
surface.\cite{brag99,bc,sopaper}

 The effect of a semiconductor surface on the
	light absorption in indirect band  semiconductor nanocrystals has
been studied in Ref. \onlinecite{brag99} and was found to be
significant. The 
 effect of the surface modeled by GBC
on the quantum confined levels  (QCLs) of an electron in zero and  weak external magnetic
fields has been studied \cite{sopaper} and the appropriate surface parameter for the CdSe spherical NCs has been determined by fitting the experimental data \cite{gupta02,sopaper}. It was shown that  the surface induces an additional 
spin--orbit splitting of  the electron states. In small CdSe NCs the respective splitting of the $1P_{e}$ states was calculated to be about 10 meV. However, in the  NCs made from the narrow gap semiconductors such as PbS and PbSe  one may expect this  effect to be larger. Indeed, a recent comparison of the absorption spectra of bare PbS NCs and PbS NCs passivated by Cl shows a significant shift of the second absorption peaks.\cite{BrittmanPCL2019}

In the present paper we study the effect of the NC surface on the electron and hole energy levels in spherical PbS and  PbSe  NCs. We derive the GBC for the four component envelope function at the abrupt spherical surface modeled as the impenetrable barrier for the carriers.  The use of the GBC for the electronic structure calculations leads to the spin orbit splitting of the electron and hole excited state energy levels.  Finally, we find GBCs that allow us to describe the difference between absorption spectra in bare PbS  and PbSe NCs and PbS NCs passivated by Cl. 
 
The paper is organized as follows. In Sec. \ref{theory}, we introduce the four-band Hamiltonians. In Sec. \ref{spherbc}, we derive the generalized boundary conditions (GBCs) for the four-band envelope functions at an abrupt surface. The effects of the surface on the electron and hole energy levels are analyzed in Sec. \ref{spherso}. In Sec. \ref{fullmodel}, we derive analytical equations for the electron and hole energy levels within the GBC model. Using these equations, we determine the phenomenological surface parameters that provide a quantitative description of the size-dependent absorption spectra of bare PbS and PbSe NCs, as well as Cl-passivated PbS NCs (Sec. \ref{absorption}). Finally, the results are summarized and discussed in Sec. \ref{sum}.

\section{\label{theory}Four band $\bm k \cdot \bm p$ model  describing carrier energy spectra   near  $L$ point of  Brillouin zone}

The PbS and PbSe  are the direct band gap semiconductors. The extremes of the conduction and valence band are at four equivalent $L$ points in the Brillouin zone. 
The energy band structure  near the
$L$ point  can be well described within the
four--band $\mathbf{k \cdot p}$  model.\cite{dimmock64,wallis66}. This model takes into account the direct interaction between the nearest conduction and valence bands and the contribution of the remote bands into the electron and hole effective masses. Following Ref. \cite{kang_wise}, we expand  the  full wave function in the homogeneous semiconductor inside the NC as:
\begin{eqnarray}
{\Phi (\mbox{\boldmath $ r$})} &=& \sum_{\mu = \pm 1/2}
\Psi^{c}_\mu(\mbox{\boldmath $ r$}) |L_{6,\mu}^{-}\rangle
 \, + \sum_{\mu = \pm 1/2}\Psi^{v}_\mu(\mbox{\boldmath $ r$}) |L_{6,\mu}^{+}\rangle
 \,
,\label{psi}
\end{eqnarray}
where $|L_{6,1/2}^{-}\rangle $ and  $|L_{6,-1/2}^{-}\rangle $ are the Bloch functions of the conduction band edge at the
$\L$--point of Brillouin zone (the upper sign "-" reflects the oddness in respect to the operation of spatial inversion); and $|L_{6,1/2}^{+}\rangle $ and  $|L_{6,-1/2}^{+}\rangle $  are the Bloch function of the valence band edge at the $\L$--point of
Brillouin zone (the upper sign "+" reflects the invariance in respect to the operation of spatial inversion). The smooth functions $\Psi^{c}_{\pm
1/2}(\mbox{\boldmath $r$})$ and $\Psi^{v}_{\pm
1/2}(\mbox{\boldmath $r$})$ are the components of the conduction band and valence band
spinor envelope functions, respectively:
\begin{eqnarray}
\Psi^c=\left( \begin{array}{c} \Psi^c_{1/2}
\\ \Psi^c_{-1/2} \end{array} \right) \, , \qquad \Psi^v=\left( \begin{array}{c} \Psi^v_{1/2}
\\ \Psi^v_{-1/2} \end{array} \right).
\end{eqnarray} 
 The bi--spinor 
envelope function $\Psi=\left( \begin{array}{c} \Psi^c
\\ \Psi^v \end{array} \right)$ is the solution of the Schr\"{o}dinger equation 
$\hat H(\hat {\bm p}) \Psi = E \Psi$, where $ \mbox{\boldmath $\hat  p$}= \hbar \hat {\bm k}= - i \hbar \mbox{\boldmath $\nabla$}$ is the momentum operator,  and the Hamiltonian $\hat H( {\bm p})$ of Ref. \cite{kang_wise} can be written in the compact form as
\begin{eqnarray}
    \hat{H}( \mbox{\boldmath $  p$})= 
    \left(  \begin{array}{cc} \left(\dfrac{E_g}{2}+ \dfrac{p^{2}_z}{2m_l^{-}}  +\dfrac{p^{2}_t}{2m_t^{-}}  \right)  \hat{U}_2 &
     \dfrac{P_l}{m_0} p_z \sigma_z +  \dfrac{P_t}{m_0} \left( p_t \sigma_t\right)   \\
 \dfrac{P_l}{m_0} p_z \sigma_z +  \dfrac{P_t}{m_0} \left( p_t \sigma_t\right)  &
- \left(\dfrac{E_g}{2}+ \dfrac{p^{2}_z}{2m_l^{+}}  +\dfrac{p^{2}_t}{2m_t^{+}}  \right)   \hat{U}_2 
\end{array} \right)
 \, \,  .  \label{hfull} 
\end{eqnarray}
 In Eq. (\ref{hfull}) the $2\times 2$ unit matrix, $\hat U_2$, and  the Pauli matrices,
$ \hat {\bm \sigma}= \{ \hat \sigma_x ,
\hat \sigma_y ,  \hat \sigma_z \}$   are  acting on the spinor components of the wave
functions ($\mu= \pm 1/2$) and the $z$ axis is directed along the $\langle 111\rangle$ direction of the cubic lattice; $E_g$ is the bulk band gap energy,  the energy $E$ is measured from the middle of the band gap, $m_0$ is the free electron mass, $p_t^2=p_x^2+p_y^2$, and  $( p_t \sigma_t)=p_x \sigma_x + p_y \sigma_y$;  $P_t$ and $P_l$ (the same as in Ref. \onlinecite{kang_wise}) represent the transverse and longitudinal momentum matrix elements taken between the conduction band edge and valence band edge Bloch functions,  and parameters $m_t^\pm$ and $m_l^\pm$ (the same as in Ref. \cite{kang_wise}) take into  account the contribution  of remote bands   to  the transverse and  longitudinal band edge effective masses, respectively.  

It was shown in Ref. \cite{kang_wise}, that anisotropy of the band structures of PbS and PbSe  is small enough to be treated by a perturbation method.  Unfortunately in  PbTe the anisotropy of the energy spectra  is very large and  the perturbation approach cannot be applied (see Table \ref{tab1}, which shows the energy band parameters for PbS, PbSe and PbTe).
Following Ref.
\onlinecite{kang_wise}, we present the Hamiltonian (\ref{hfull}) as $\hat H = \hat H_0 + \hat H_{an}$,
where the spherically symmetric part $\hat H_0$  can be written as\cite{kang_wise} 
\begin{eqnarray}
\hat H_0( \mbox{\boldmath $  p$})= 
    \left(  \begin{array}{cc} \left(\dfrac{E_g}{2} + \dfrac{\alpha_cp^2}{2 m_0} \right)   \hat U_2 &
     \dfrac{ P}{m_0} ({\bm p}{\bm \sigma}) \\ \dfrac{P}{m_0} ({\bm p}{\bm \sigma})
  &  -\left(\dfrac{E_g}{2} + \dfrac{\alpha_v p^2}{2 m_0} \right)     \hat U_2 
\end{array} \right)
\label{h0}
\end{eqnarray}
and the anisotropic perturbation $\hat H_{an}$ can be found in appendix of Ref. \onlinecite{kang_wise}. The parameters $\alpha_c=m_0/m^-$, $\alpha_v=m_0/m^+$ and $P$ are  chosen such that the effect of perturbation is minimized \cite{kang_wise}: $
3P^2=2P_t^2+P_l^2$, $3/m^\pm = 2/m_t^\pm + 1/m_l^\pm$. For $\alpha_c > 0$, $\alpha_v > 0$ (typical for PbSe and PbS, see Table \ref{tab1}),  Hamiltonian $\hat H_0$ in  Eq. (\ref{h0}) describes the spherically symmetric conduction and valence bands at small real values of the wave vector $k \ll 2\pi/a_0$, where $a_0$ is the lattice constant. The respective dispersion laws $E_c(k)$ and $E_v(k)$ are the solutions of the energy  dispersion equation, where the energy $E$ is measured from the middle of the gap.:
\begin{equation}
\left( \frac{\hbar^2 k^2}{2 m_0} \alpha_c + \frac{E_g}{2} -  E
\right) \left( \frac{\hbar^2 k^2}{2 m_0} \alpha_v  +\frac{E_g}{2}+
E \right) + \frac{\hbar^2 k^2}{2 m_0} E_p =0~.
\label{dispersion}
\end{equation}
This yields the electron and hole energy dispersion relations, $E_{c,v}(k)$
\cite{kang_wise}
 \begin{equation}
E_{c,v}(k)=\frac{\alpha^-k^2}{2} \pm \frac{1}{2}\sqrt{(E_g + \alpha^+ k^2)^2 +2E_p(\hbar k)^2/m_0} \, ,
\label{disp}
\end{equation}
where the signs $\pm$ are related to the conduction /valence  band respectively and parameters
 \begin{equation}
\alpha^+= \frac{\hbar^2}{2m_0}(\alpha_c + \alpha_v) \, , \quad \alpha^-= \frac{\hbar^2}{2m_0}(\alpha_c - \alpha_v) \, . 
\label{disp1}
\end{equation}


 Near the symmetry point $k=0$, the parabolic approximation to the energy dispersion law reads $E_{c,v}= \hbar^2k^2/2m_{c,v}^*$, where the band edge masses $m_c^*$ and $m_v^*$ in spherical approximation can then be found as ${m_0}/{m^{*}_{c,v}} = \alpha_{c,v} + {E_p}/{E_g} $, 
One can see from Table \ref{tab1} 
\begin{table}[h!]
\begin{center}
\caption{\label{tab1} Parameters of the four band model of PbS,  PbSe, and PbTe \cite{dimmock64} taken from  (a) Ref. \onlinecite{kang_wise} and (b)  Ref. \onlinecite{andreev99} and calculated values of the band edge effective masses. }
\begin{tabular}{cccccc}
\hline
Parameters & PbS$ ^a$ & PbS$ ^b$ & PbSe$ ^a$ &  PbSe$ ^b$  & ~PbTe
\\
\hline 
$E_g$ ($T=300$ K) (eV) & 0.41 &  & 0.28 & & ~0.31\\
$E_g$ ($T=12$ K) (eV) &  & 0.299 &  & 0.166 &~ 0.190 \\
$m_0/m_t^{-}$ &1.9  & 1.3 & 4.3  &  4.3 & ~13.5 \\
$m_0/m_l^{-}$& 3.7 & 2.9 & 3.1  & 3.1 & ~1.8\\
$\alpha_c$ & 2.5 &  & 3.9  & 3.9 & ~9.6 \\
$m_0/m_t^{+}$ & 2.7 & 1.6& 8.7 & 8.7   & ~15.0 \\
$m_0/m_l^{+}$ & 3.7 & 2.9 & 3.3 & 3.3 & ~1.8 \\
$\alpha_v$ & 3.0 &  & 6.9  & 6.9 & ~ 10.6 \\
$2P_t^2/m_0$ (eV) & 3.0 & 1.4  & 3.0  & 1.7 & ~6.1 \\
$2P_l^2/m_0$ (eV) & 1.6 & 3.3 & 1.7 & 3.0 & ~1.5 \\
$E_p$ (eV) & 2.5 &  & 2.64  & 2.22 &~ 4.57 \\
$a^*=\hbar/2P$\,(nm)& 0.124&&0.120&& \\
$m_c^*/m_0$ ($T=300$ K) & & &0.075  & 0.085 & ~0.055\\
$m_c^*/m_0$ ($T=12$ K) & & &0.051 &0.058 & ~0.024 \\
$m_v^*/m_0$ ($T=300$ K) & & &0.061 & 0.067 &~ 0.045 \\
$m_v^*/m_0$ ($T=12$ K) & & &0.044 &0.049 & ~0.021 \\
\hline
\end{tabular}
\end{center}
\end{table}
that the typical values of $\alpha_{c,v}$ are between 3 and 7, and the typical value of $E_p/E_g$ is about 10. The main contribution to the band edge masses comes from the interaction between the nearest conduction and valence bands. That is because of the small value of the band gap energies in these materials. The first consequence of this fact is that one may expect the strong temperature dependence of the band edge effective masses via the temperature dependence of the band gap energy. We also may expect the large effect of the dispersion nonparabolicity caused  by the energy dependence of the conduction band electron, $m_c(E)$, and valence band electron (hole), $m_v(E)$,  effective masses. 
An approximate nonparabolic dispersion 
\begin{equation}
	E_{c,v} =  \pm \frac{E_g}{2} \pm \frac{\hbar^2k^2}{ 2 m_{c,v}(E_{c,v})} \, , \qquad	{\rm where}~~~{m_0\over m_{c,v}(E)}=\alpha_{c,v}+{E_p\over E_g/2\pm E},
		\label{eq:9}
		\end {equation}	
		can be obtained from Eq. (\ref{disp})  by using $\alpha_v=0$ for electrons or $\alpha_c=0$ for holes, respectively.  
 
\begin{figure}[H]
\centering
\includegraphics[width=14cm]{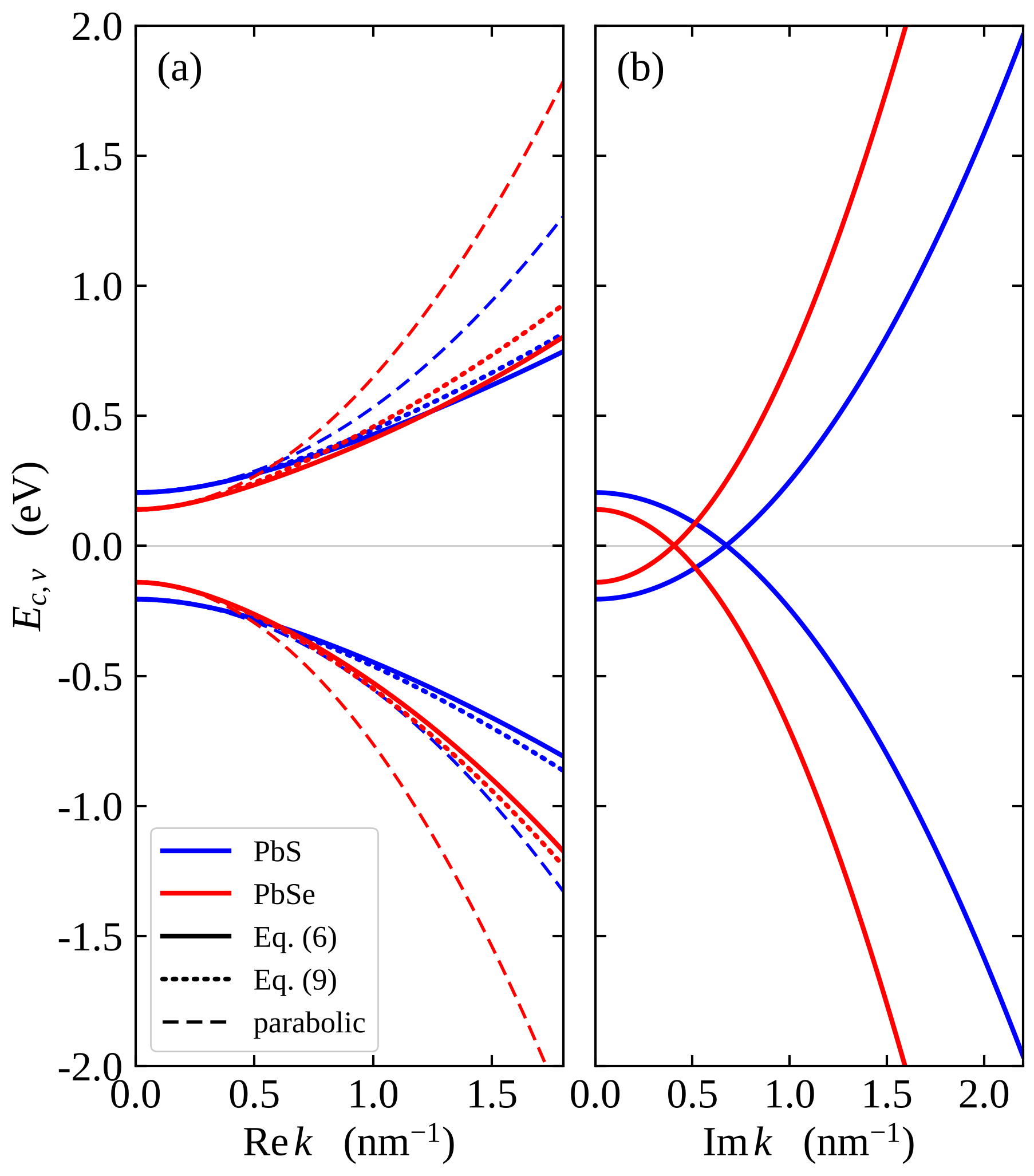}
\caption{ (a) The bulk dispersion of electrons and holes (in the electron representation) in PbS   (blue lines) and  PbSe (red lines) calculated in exact spherical approximation using Eq. (\ref{disp}). Dotted lines show the dispersion calculated  using  approximate  expression of Eq. (\ref{eq:9}).   For  comparison we show  the parabolic dispersion (dashed lines) calculated for  electron and hole effective masses at the band edge of the conduction and valence bands, correspondingly. (b) The  imaginary  values of momenta, defined in Eq. (\ref{eq:20}) that satisfy the condition $E_{c,v}(k)=E$ for any arbitrary energy $E$ in PbS   (blue lines) and  PbSe (red lines). The real part of momentum  that satisfy the condition $E_{c,v}(k)=E$ can be extracted from the carrier dispersion shown in panel (a). 
}
\label{ek}
 \end{figure}
Figure \ref{ek}(a) shows the dispersion  of electron and hole (in the electron representation) calculated using Eq. (\ref{disp}). One can see, for example, that $m_c(E_g)$ is by about factor 1.5 larger than the band edge mass $m_c^*=m_c(E_g/2)$, and $m_v(-E_g)$ is by about factor 1.5 larger than the band edge mass $m_v^*=m_v(-E_g/2)$ resulting in the strong nonparabolicity of the energy dispersion. Therefore, the parabolic  model with constant effective masses can not describe the QCLs with energies larger than band gap satisfactory\cite{kang_wise}.  On the other hand, the approximate nonparabolic dispersion laws of Eq. (\ref{eq:9}) work well in a much larger energy regions.  If the energy $E$ of an electron level lies below the conduction-band minimum, the condition $E_c(k)=E$ can be satisfied only for $k^2<0$. Conversely, if the energy $E$ of a hole level lies above the valence-band maximum, the condition $E_v(k)=E$ can also be satisfied only for $k^2<0$. Thus, in both cases, the corresponding wave vectors $k$ are purely imaginary. Figure \ref{ek}(b) shows the dependence of the corresponding energies on the imaginary wave vector ${\rm Im}k$.

\section{\label{spherbc} General boundary conditions for the envelope wave functions}

\subsection{Envelope flux density matrix}

We will consider semiconductor NCs whose surface can be
modeled by an impenetrable barrier.  To find unperturbed envelope
wave functions that are described by the Hamiltonian of  Eq.
(\ref{h0}) in the bulk region one should impose appropriate boundary conditions on these functions at the NC surface.
For the impenetrable barrier the general boundary conditions (GBCs)
have been shown in Ref. \onlinecite{bc}  to guarantee the vanishing
of the normal to the surface component of the envelope flux density
matrix. In a spherical NC the normal envelope flux density
matrix, $J_{\tau}^{\lambda  \, \eta}(\mbox{\boldmath $r$})$,  should
vanish at any point of the NC surface:\cite{sopaper}
\begin{eqnarray}
J_{\tau}^{\lambda  \, \eta}(\mbox{\boldmath $r$}) {\left.
\right|_{r=a}} =  
\left. \frac{1}{2}  \left[ \left( {\Psi}_\lambda ,
 \hat  V_{\bm \tau}
{\Psi}_\eta \right) + \left(  \hat  V_{\bm \tau}
{\Psi}_\lambda  , {\Psi}_\eta \right) \right] \right|_{r=a} = 0 \, ,
\label{generalflux}
\end{eqnarray}
where $a$ is the NC radius,  $\mbox{\boldmath
$\tau$}=\mbox{\boldmath $ r$}/r$ (in spherical coordinates
$\mbox{\boldmath $r$}$ is defined as $\mbox{\boldmath
$r$}\equiv(r,\Theta,\varphi )$), and $\hat V_{\bm \tau}= {\bm \tau} \hat {\bm V}$ is the normal projection of the envelope velocity operator $\hat V=\partial \hat H^0({\bm p})/\partial
\mbox{\boldmath $p$} $.   This general requirement of Eq.
(\ref{generalflux}) should be satisfied for two arbitrarily chosen
eigenfunctions ${ \Psi}_\lambda $ and ${ \Psi}_\eta$   of the
Hamiltonian $\hat H_0$ defined in Eq. (\ref{h0})  with energies
$E_\lambda$ and $E_\eta$, respectively.  Substituting the explicit
expressions for the normal projection of the envelope  velocity operator
\begin{eqnarray}
\hat V_{\bm \tau} = \frac{1}{m_0}
    \left(  \begin{array}{cc} \alpha_c ({\bm p} {\bm \tau})\hat U_2 &
      P ({ \bm \sigma}{\bm \tau}) \\ P ({\bm \sigma}{\bm \tau})
  &    - \alpha_v ({\bm p}{\bm \tau})\hat U_2 
\end{array} \right)
\label{vtau}
\end{eqnarray}
 into Eq. (\ref{generalflux}) one  obtains:
\begin{eqnarray}
&&J_{\tau}^{\lambda  \, \eta} (\mbox{\boldmath $r$})
\left|_{r=a}\right.  =
 \frac{i  \hbar}{2m_0} \left[ \alpha_c \left(\Psi^c_\eta
\mbox{\boldmath $\tau$} \cdot \mbox{\boldmath $\nabla$} \Psi^{c
*}_{\lambda } - \Psi^{c *}_{\lambda } \mbox{\boldmath $\tau$} \cdot
\mbox{\boldmath $\nabla$} \Psi^c_\eta \right) \right.\nonumber \\&-&  
\left. \left. \alpha_v \left(\Psi^v_\eta
\mbox{\boldmath $\tau$} \cdot \mbox{\boldmath $\nabla$} \Psi^{v
*}_{\lambda } - \Psi^{v *}_{\lambda } \mbox{\boldmath $\tau$} \cdot
\mbox{\boldmath $\nabla$} \Psi^v_\eta \right) + 
\frac{2 P }{i \hbar}\left(  \Psi^{c *}_{\lambda }  ({\bm \sigma}{\bm \tau}){\Psi}^{v}_\eta    +
 { \Psi}^{v *}_{\lambda} ({\bm \sigma}{\bm \tau})
\Psi^{c}_\eta  \right) \right] \right|_{r=a} =0 \, . \label{fluxelec}
\end{eqnarray}

For qualitative understanding  of  GBC effects we will start from analyses of a simplified model in which $\alpha_c \alpha_v = 0$, which results in the approximate nonparabolic dispersion for electrons and holes discussed in the previous Section.   
\subsection{A simplified model with  $\alpha_c \alpha_v =0$}

Let us first consider the case $\alpha_v=0$ and $\alpha_c \ne 0$. In this case the Eq. (\ref{fluxelec}) is satisfied for all arbitrary chosen energy states
$\lambda$ and $\eta$ if and only if the boundary conditions for
$\Psi_\lambda$ as well as for $\Psi_\eta$  are given by:
\begin{eqnarray}
\Psi^c(r=a,\Theta ,\varphi)  = \left. T_c a^* \left( \alpha_c
\frac{\partial \Psi^c(r,\Theta ,\varphi)}{\partial r} -
\frac{2P}{\hbar i} ({\bm \sigma}{\bm \tau})\Psi^v(r,\Theta ,\varphi) \right)
\right|_{r=a}  \, , \label{sph_cBC}
\end{eqnarray}
 where $T_c$ is a real number constant independent of $E$ and  $a^*=\hbar/2P$ is the characteristic length constant of the interband interaction. 
 We assume that the surface of the NC
also possesses the spherical symmetry and is characterized by the
same GBCs at any point of the surface. In this case the surface parameter $T_c$  in
the  GBCs of Eq. (\ref{sph_cBC}) is unique characteristic  of the NC surface  and it is independent of angles $\Theta$ and
$\varphi $.

Importantly, in the case $\alpha_c \alpha_v =0$, the wave function components $\Psi^{c}$  and  $\Psi^{v}$ are not independent. For $\alpha_v =0$, it is convenient to express the
components  of the spinor ${\Psi}^v$ through the components of the spinor 
$\left( {\bm \sigma} {\bm \nabla} \right)\Psi^c$ using the second row of  Eq. (\ref{h0}):  
\begin{eqnarray}
{\Psi}^v = \frac{i\hbar}{2P} \left( \alpha_c - \frac{m_0}{m_c(E)}
 \right)  \left( {\bm \sigma} {\bm \nabla} \right)\Psi^c \, ,
\label{psiv}
\end{eqnarray}
where the energy dependent electron effective mass $m_c(E)$ is defined in Eqs. (\ref{eq:9}) (note, that the energy $E$ is calculated from the middle of the band gap). Then, the product  $({\bm \sigma}
{\bm \tau})\Psi^v$ in Eq. (\ref{sph_cBC}) can  be written as:
\begin{eqnarray}
({\bm \sigma} {\bm \tau})\Psi^v &=&
\frac{i\hbar}{2P} \left( \alpha_c - \frac{m_0}{m_c(E)}  \right)
\frac{\partial \Psi^c}{\partial r}\nonumber \\
&+& \frac{i \hbar}{2 P r} \Delta m_c(E) (\mbox{\boldmath $\hat \sigma$}
 \mbox{\boldmath $\hat  L$}) \Psi^c \, ,
\label{psivtau}
\end{eqnarray}
where $\Delta m_c(E) = m_0/m_c(E)-\alpha_c =E_p/(E_g/2+E) $. Substituting Eq. (\ref{psivtau})
into Eq. (\ref{sph_cBC}) one obtains the  general expression for 
the conduction band component of the envelope wave function at the
NC surface:
\begin{eqnarray}
\Psi^c(r=a,\Theta ,\varphi)  = \gamma_c \left( \frac{m_0}{m_c(E)}
\frac{\partial \Psi^c(r,\Theta ,\varphi)}{\partial r} \right.
\nonumber \\ \left. \left.  - \frac{1}{r}\Delta m_c(E) (\mbox{\boldmath
$\hat \sigma \hat  L$})\Psi^c(r,\Theta ,\varphi) \right)
\right|_{r=a}  \, , \label{sph_ccBC}
\end{eqnarray}
where $\gamma_c=T_c a^*$.

Similarly,  in the case $\alpha_c=0$ and  $\alpha_v\ne 0$, the boundary conditions for
$\Psi_\lambda$ as well as for $\Psi_\eta$  should be given by:
\begin{eqnarray}
\Psi^v(r=a,\Theta ,\varphi)  = \left. T_v a^* \left( \alpha_v
\frac{\partial \Psi^v(r,\Theta ,\varphi)}{\partial r} +
\frac{2P}{\hbar i} ({\bm \sigma}{\bm \tau})\Psi^c(r,\Theta ,\varphi) \right)
\right|_{r=a}  \, , \label{sph_vBC}
\end{eqnarray}
 where $T_v$ is  real number constant,   unique characteristic  of the NC surface  and it is independent of $E$ and of angles $\Theta$ and
$\varphi $.
Expressing ${\Psi}^c$ through
$\mbox{\boldmath $\nabla$} \Psi^v$ and substituting into Eq. (\ref{sph_vBC}) one obtains the  general expression for the conduction band component of the envelope wave function at the
NC surface:
\begin{eqnarray}
\Psi^v(r=a,\Theta ,\varphi)  = \gamma_v \left( \frac{m_0}{m_v(E)}
\frac{\partial \Psi^v(r,\Theta ,\varphi)}{\partial r} \right.
\nonumber \\ \left. \left.  - \frac{1}{r}\Delta m_v(E) (\mbox{\boldmath
$\hat \sigma \hat  L$})\Psi^v(r,\Theta ,\varphi) \right)
\right|_{r=a}  \, , \label{sph_vvBC}
\end{eqnarray}
where $\gamma_v=T_v a^*$, $m_v(E)$ is defined in Eqs. (\ref{eq:9}) and $\Delta m_v(E) = m_0/m_v(E)-\alpha_v=E_p/(E_g/2-E)$. 

In the case $\alpha_c=\alpha_v=0$ both Eq. (\ref{sph_cBC}) and Eq. (\ref{sph_vBC}) can be written and should be equivalent. This gives the connection between parameters $\gamma_c=T_c a^*$ and $\gamma_v=T_v a^*$ as $\gamma_c \gamma_v = a^{*2}$ or $T_c T_v=1$. One can see, that in the case $\alpha_c \alpha_v = 0$, the conduction band and valence band wave function spinors can not satisfy the standard BC $\Psi^c=0$ and $\Psi^v=0$ at the surface simultaneously. Importantly, only one BC, either of  Eq. (\ref{sph_cBC}) or of Eq. (\ref{sph_vBC}) should be used for calculating the electron and hole QCLs in the approximate model $\alpha_c \alpha_v=0$. In other words, the self-adjoint extension of the Hamiltonian $\hat H_0$ and the whole energy spectrum in NC are controlled by only one surface parameter that can be chosen as $T_c$ or $T_v$ ($\gamma_c$ or $\gamma_v$).

\subsection{Complete model $\alpha_c \alpha_v \ne 0$}

In a case $\alpha_c \alpha_v \ne 0$, one can apply the SBC: $\Psi^c=0$ and $\Psi^v=0$ at the surface as it was done in Ref. \onlinecite{kang_wise}. However, it is easy to see that the SBC in this case represent only one realization of the GBC.  In a general case,
 the Eq. (\ref{fluxelec}) is satisfied for all arbitrary chosen energy states
$\lambda$ and $\eta$ if and only if the boundary conditions for
$\Psi_\lambda$ as well as for $\Psi_\eta$  are given by:
\begin{eqnarray}
\Psi = \frac{i\, m_0 }{2P}
  \hat {T}_{sur}^{-1} \hat  V_{\bm \tau} \Psi \,
\label{sphBC}
\label{gbc1}
\end{eqnarray}
where $\hat V_{\bm \tau}$ is given by Eq. (\ref{vtau}), $\hat T_{sur}^{-1}$ denotes the reverse matrix to $\hat T_{sur}$, and $\hat T_{sur}$ is a Hermitian surface matrix:
\begin{eqnarray}
 \hat T_{sur} = \left( \begin{array}{cc}
t_c &  -i t ({\bm \sigma}{\bm \tau}) \\
 it ({\bm \sigma}{\bm \tau}) & - t_v \end{array} \right) \, ,
\label{gbc2}
\end{eqnarray}
 where $t_c$, $t_v$ and $t$ are real number constants independent of $E$ and of angles $\Theta$ and $\varphi $.

The GBCs of Eqs. (\ref{gbc1},\ref{gbc2}) as well as the SBC $\Psi=0$ can be satisfied only if 
the envelope wave function of a state with energy $E$ is
written as a linear superposition of two plane wave eigenfunctions
of the Hamiltonian, Eq.(\ref{h0}):  ${\Psi}  = C_1{\Psi}_1(k_1)
+ C_2{ \Psi}_2(k_2)$.  Here  $k_{1,2}^{2}$ are the solutions of the
bulk energy  dispersion equation Eq. \ref{dispersion}, that can  be written as 
\begin{equation}
k_{1,2}^2=\frac{m_0}{\hbar^2\alpha_c\alpha_v}\left[-B(E)\pm\sqrt{B^2(E)+4\alpha_c\alpha_v\left(E^2-E_g^2/4\right)}\right].
\label{eq:20}
\end{equation}
where $B(E)=E_p+ E_g\alpha^+/2+E\alpha^-.$   The $k_{1,2}^2 $ are either  positive or negative and corresponding   $k_{1,2}$  are real or imaginary.  The energy dependencies of $E_{c,v}$  on  real and imaginary parts of $k_{1,2}$  are shown in  Fig.\ref{ek}. Note, that two solutions $k_{1,2}^2$ exist only in general case $\alpha_c\alpha_v \ne 0$. In the approximate case $\alpha_c\alpha_v = 0$ the dispersion equation Eq. \ref{dispersion} does not contain the $k^4$ term. 

The Eq. (\ref{dispersion}) is the same as the one, describing the energy dispersion of the electrons and light holes near the $\Gamma$ point in zinc blend semiconductors with zero spin--orbit splitting of the valence band \cite{efros_prb98,bcpaper}. However, the characteristics parameters of the PbSe, PbS are different from those for zinc blend semiconductors. For the latter, one is usually interested in the states whose energy is either on
the order of or smaller than the semiconductor energy gap, so that
the condition  $E_p/|\alpha_c \alpha_v| \gg  E_g, |E \pm E_g/2|$ takes place. 
For typical values of the bulk energy band parameters of zinc blende semiconductors,  the second
wave vector $|k_2| \approx \sqrt{-2E_pm_0/\hbar^2\alpha_c \alpha_v}\sim 1\,  \AA^{-1}$  lies beyond the first
Brillouin zone in which the multiband effective mass approximation is valid.  The corresponding
eigenvector, ${\Psi}_2(k_2)$ is a spurious  solution (either
evanescent or propagating, depending on the sign of the product
$(\alpha_c \alpha_v)$ \cite{bc}  and  must be excluded from the
expansion of the total wave function. In this case, the  smooth envelope $\Psi_1(k_1)$ can not satisfy SBC or GBC of Eq. (\ref{gbc1},\ref{gbc2}). In this case, the approximate procedure of deriving the truncated GBC developed in Ref. \cite{bcpaper} should be used. However, for the characteristics parameters of PbS and PbSe the absolute values of both roots are of the same order and the problem of spurious solution does not arise. Indeed, the condition  $E_p/|\alpha_c \alpha_v| \gg  E_g, |E\pm E_g/2|$ is not the case for PbS and PbSe if one uses the real values of $\alpha_c$ and $\alpha_v$. 

Thus, in  the case $\alpha_c \alpha_v \ne 0$  GBC of Eq. (\ref{gbc1},\ref{gbc2}) contain 3 unknown real boundary parameters. We will look now how to connect the parameters $t_c$, $t_v$ and $t$ with $\gamma_c$ and $\gamma_v$ introduced for the case $\alpha_c \alpha_v =0$, so that the GBC of Eq. (\ref{gbc1},\ref{gbc2}) transforms smoothly to the boundary conditions of Eq. (\ref{sph_cBC}) or Eq. (\ref{sph_vBC}) respectively. We consider the case $\alpha_c \alpha_v \ge 0$ corresponding to the parameters listed in Table \ref{tab1} and find:
\begin{eqnarray}
t_c(\alpha_c,\alpha_v) = \frac{a^*}{\gamma_c} \frac{\alpha_c}{\alpha_c+\alpha_v} \, \nonumber \\
t_v(\alpha_c,\alpha_v) = \frac{a^*}{\gamma_v} \frac{\alpha_v}{\alpha_c+\alpha_v} \,  \\
t(\alpha_c,\alpha_v) = \frac{1}{2} \frac{\alpha_c-\alpha_v}{\alpha_c+\alpha_v} \,  \, , \nonumber 
\end{eqnarray}
so that $t_c(0,\alpha_v \ne 0)=t_v(\alpha_c \ne 0,0)=0$, $t_c(\alpha_c \ne 0,0)=1/T_c$, $t_v(0,\alpha_v \ne 0)=1/T_v$, $t_c(0,0)=1/2T_c$, $t_v(0,0)=1/2T_v$, $t(\alpha_c \ne 0,0)=-t_v(0,\alpha_v \ne 0)=1/2$, and $t(0,0)=0$. Using this parameters one can rewrite the GBC of Eq. (\ref{gbc1},\ref{gbc2}) as following:
\begin{eqnarray}
\Psi^c \left|_{r=a} \right. = \left.  \gamma_c \left( \left(\alpha_c+ \alpha_v\right)
\frac{\partial \Psi^c}{\partial r} -
\frac{2P}{\hbar i} ({\bm \sigma}{\bm \tau})\Psi^v \right)
\right|_{r=a}  \, , \label{sph_cGBC} \nonumber\\
\Psi^v\left|_{r=a} \right. = \left. \gamma_v \left( \left(\alpha_c+ \alpha_v\right)
\frac{\partial \Psi^v}{\partial r} +
\frac{2P}{\hbar i} ({\bm \sigma}{\bm \tau})\Psi^c \right)
\right|_{r=a}  \, . \label{sph_vGBC}
\end{eqnarray}
 The conditions of Eq. (\ref{sph_vGBC}) can be satisfied simultaneously in general case $\alpha_c \alpha_v > 0$ if $\Psi = C_1 \Psi_1(k_1) + C_2 \Psi_2(k_2)$ with any real numbers $T_c=\gamma_c/a^*$ and $T_v=\gamma_v/a^*$, including the case of SBC with $T_c=T_v=0$. However, in the case $\alpha_c \alpha_v =0$, the constrain  $\gamma_c \gamma_v =a^{*2}$ should be used.

\section{\label{spherso}Electron and hole energy levels:
 the  surface induced spin--orbit splitting}

\subsection{Symmetry of the eigenfunctions}

The symmetry of the Hamiltonian $\hat H_0$ is the same as for the Dirac electron in a central potential \cite{LL}. Therefore, the eigenfunctions $\Psi$ are the simultaneous eigenstates of the total angular momentum operator $\hat {\bm J}$ and parity operator $\hat \Pi $ \cite{kang_wise}, where
\begin{eqnarray}
\hat {\bm J} = \hat {\bm L} + \frac{\hbar}{2}\hat {\bm \Sigma} \, , \qquad 
\hat {\bm \Sigma} = \left( \begin{array}{cc} {\bm \sigma} & 0 \\ 0 & {\bm \sigma} \end{array} \right) \, ,\nonumber \\
\hat \Pi = \left( \begin{array}{cc} - \hat {\bm P } \otimes \hat U_2 & 0 \\ 0  & \hat {\bm P} \otimes \hat U_2 \end{array} \right) \, ,
\end{eqnarray}
where $\hat {\bm P}$ is the usual spatial inversion operator. The detailed expressions for the eigenfunctions labeled by the total angular momentum $j$, its projection $m$ and by the parity $\pi = (-1)^{j\pm 1/2}=\pm 1$ can be found in Ref. \onlinecite{kang_wise}. We rewrite here the eigenfunctions  in the more convenient for the following consideration compact form:
\begin{eqnarray}
\Psi_{\pi,j,m}({\bm r}) = \left( \begin{array}{c}
 f_l(r) \Omega_{l+1/2,l,m} \\   f_\lambda(r)  \Omega_{\lambda-1/2,\lambda,m} \end{array} \right) \, , \quad j=l+1/2=\lambda-1/2,\, \, \pi =(-1)^{j+1/2}, \nonumber\\
 \Psi_{\pi,j,m}({\bm r}) = \left( \begin{array}{c}
 g_l(r) \Omega_{l-1/2,l,m} \\ - g_\lambda(r)  \Omega_{\lambda+1/2,\lambda,m} \end{array} \right) \, , \quad j=l-1/2=\lambda+1/2,\, \, \pi =(-1)^{j-1/2}, \label{eq:29}
\end{eqnarray}
where $\Omega_{j,j\pm 1/2,m}$ are the spherical spinors defined as in  Ref. \onlinecite{ll} (note, that for consistency one has to use the  definition of the spherical harmonics as in Ref. \onlinecite{ll}, which is different from those used in Ref. \onlinecite{kang_wise}), $l=j \pm 1/2$ is the orbital angular momentum determining the symmetry of the conduction band spinor envelope function $\Psi^c$ and $\lambda=2j-l=j\mp 1/2$ is the orbital angular momentum determining the symmetry of the valence band spinor envelope function $\Psi^v$.  The radial functions $f_{l,\lambda}(r)$ and $g_{l,\lambda}(r)$ satisfy the systems of the radial equations :\cite{kang_wise}
\begin{eqnarray}
\left[ \epsilon_g/2 - \epsilon -\alpha_c  \Delta_{l} \right] f_l(r) - \upsilon A^{-}_{\lambda} f_\lambda(r) = 0 \, , \nonumber \\
- \upsilon A^{+}_l f_l(r) - \left[ \epsilon_g/2 + \epsilon -\alpha_v  \Delta_{\lambda} \right] f_\lambda(r) = 0 \, ,
\label{eqf}
\end{eqnarray}  
where $\lambda=l+1$, and 
\begin{eqnarray}
\left[ \epsilon_g/2 - \epsilon -\alpha_c  \Delta_{l} \right] g_l(r) + \upsilon A^{+}_{\lambda} g_\lambda(r) = 0 \, , \nonumber \\
+ \upsilon A^{-}_l g_l(r) - \left[ \epsilon_g/2 + \epsilon -\alpha_v  \Delta_{\lambda} \right] g_\lambda(r) = 0 \, ,
\label{eqg}
\end{eqnarray}
 where $\lambda=l-1$. In Eqs. (\ref{eqf},\ref{eqg}) we used $\epsilon_g= 2m_0E_g/\hbar^2$,  $\epsilon = 2m_0E/\hbar^2$, $\upsilon = 2P/\hbar=1/a^*$,  and
\begin{eqnarray}
\Delta_{l}=
{\partial^2\over \partial^2 r}+{2\over r}{\partial\over
\partial r}-{{l}({l}+1)\over r^2} \, , \, \, A_l^+ = - \frac{\partial}{\partial r} + \frac{l}{r} \, , \, \, A_l^- =  \frac{\partial}{\partial r} + \frac{l+1}{r} \, .
\end{eqnarray}
The radial functions $f_{l,\lambda}(r)$ and $g_{l,\lambda}(r)$ can be written as 
\begin{eqnarray}
f_{l} = \frac{C_{l1}^f}{a^{3/2}} j_{l}(k_1 r) + \frac{C_{l2}^f}{a^{3/2}} j_{l}(k_2 r)  \, ,\, \nonumber\\
f_{l+1}=  \frac{C_{l1}^f}{a^{3/2}} \Lambda(k_1)  j_{l+1}(k_1 r) +  \frac{C_{l2}^f}{a^{3/2}} \Lambda(k_2)j_{l+1}(k_2 r)  \, , \label{eq:33f} \\
g_{l} = \frac{C_{l1}^g}{a^{3/2}} j_{l}(k_1 r) + \frac{C_{l2}^g}{a^{3/2}} j_{l}(k_2 r) \, , \nonumber \\
g_{l-1}=  -\frac{C_{l1}^g}{a^{3/2}} \Lambda(k_1)  j_{l-1}(k_1 r) -  \frac{C_{l2}^g}{a^{3/2}} \Lambda(k_2)j_{l-1}(k_2 r)  \, ,\label{eq:33A}
\end{eqnarray}
where $j_l$ are spherical Bessel functions, $k_1^2$ and $k_2^2$ are the roots of Eq. (\ref{dispersion}) for a definite value of the state energy $E$ and defined in Eq. (\ref{eq:20}). They can be rewritten via dimensionless variable as:
\begin{eqnarray}
\frac{\hbar^2k_{1,2}^2}{2m_0} &=& \frac{-b\pm \sqrt{b^2 + 4 \alpha_c\alpha_v(\epsilon^2 - \epsilon_g^2/4)}}{2\alpha_c\alpha_v}   \, \quad if \quad \alpha_c \alpha_v \ne  0 \, , \nonumber\\
\frac{\hbar^2k_{1}^2}{2m_0} &= &\frac{\epsilon^2 - \epsilon_g^2/4}{b}   \, \quad if \quad \alpha_c \alpha_v  = 0 \, , 
\end{eqnarray}
 where 
$b=\upsilon^2 + \alpha_c(\epsilon_g/2 + \epsilon) + \alpha_v(\epsilon_g/2 - \epsilon)$  and 
\begin{eqnarray}
\Lambda(k_i) =\frac{\epsilon_g/2 -\epsilon + \alpha_c k_i^2}{\upsilon k_i} =  -\,\frac{\upsilon k_i}{\epsilon_g/2 +\epsilon+ \alpha_v k_i^2}
\end{eqnarray}
for $i=1,2$.  
 The energies $E$ and the coefficients $C_{l1}^{f,g}$ and  $C_{l1}^{f,g}$ are to be determined by the boundary conditions and the wave function normalization conditions
\begin{eqnarray}
\int_0^a [f_l^2(r) + f_{l+1}^2(r)] r^2 dr = 1 \, ,\nonumber \\
\int_0^a [g_l^2(r) + g_{l-1}^2(r)] r^2 dr = 1 \, . 
\end{eqnarray}

 In Ref. \onlinecite{kang_wise} the SBC $f_{l,\lambda}(a)=f_{l,\lambda}(a)=0$ have been used. They give the QCLs with energies $E>E/2$ (the conduction band electron levels) and $E< - E/2$ (the valence band electron levels). For this energies and with the condition $\alpha_c \alpha_v > 0$ that is hold for the parameters of PbS and PbSe, one of roots $k_1^2 > 0$, and the second root $k_2^2 < 0$. We do not restrict here by this case and note, that when the root $k_i^2<0$ one has to replace $j_l(i|k_i|r)=(i)^lI_l(|k_i|r)$, where $I_l$ is the modified spherical Bessel function and choose the coefficients $C_{li}^{f,g}$ to be real for the even $l$ and imaginary for the odd $l$. This allows one to choose the radial wave functions $f_{l,\lambda}$ and $g_{l,\lambda}$ to be always real.

For the SBCs and the case $\alpha_c \alpha_v > 0$ the conduction band electron and valence band electron energy levels and respective wave functions have been calculated in Ref. \onlinecite{kang_wise}. 
 For comparison, later in Fig. \ref{Fig.4} we show the size dependencies of the ground $1S_e1S_h$ optical transition calculated using SBC together with size dependence of this transition  calculated using GBCs. Within SBC model the spin--orbit splittings of the first excited states $\Delta_c = E(1P^c_{3/2}) - E(1P^c_{1/2}) <0$ and $\Delta_v = E(1P^v_{3/2}) - E(1P^v_{1/2}) >0$ are small and have different signs. The ordering of the valence band energy levels is the same as the valence band ordering in diamond--like semiconductors with nonzero spin--orbit interaction, while the ordering of the conduction electron levels is opposite. 
\subsection{Surface effects in the approximate model $\alpha_c \alpha_v = 0$}

To see how  GBCs  affect the conduction  electron energy levels  and spin--orbit splitting we first consider the approximate case $\alpha_v =0$. In this case the functions $f_l$ and $g_l$ satisfy the same radial Schr\"{o}dinger equation
\begin{eqnarray}
- \,  \frac{\hbar^2}{2m_c( E)} \Delta_{l} f_l(r) =
 \left(E - \frac{E_g}{2}\right) f_l(r) \, , f_l \leftrightarrow g_l \, . \label{Kaneel} 
\end{eqnarray}
At the same time substituting the conduction band spinor functions $\Psi^c = f_l(r) \Omega_{l+1/2,l,m}(\Theta ,\varphi )$ and  $\Psi^c = g_l(r) \Omega_{l-1/2,l,m}(\Theta ,\varphi )$ describing the states with $j_c=l +1/2$ and $j_c = l-1/2$  respectively into
 the GBCs of Eq. (\ref{sph_ccBC})   we obtain the simplified boundary conditions for the radial functions $f_l(r)$ and $g_l(r)$ at the $r=a$ spherical surface with only one surface parameter $\gamma_c=T_c a^*$: 
\begin{eqnarray}
f_l(a)= \gamma_c \left[ \frac{m_0}{m_c(E)} f_l^{ '}(a) - \frac{l}{a}\Delta m_c(E) 
f_l(a) \right] \, , \quad l=0,1,2 \dots \, , \nonumber\\ \label{bcfso}
g_l(a)= \gamma_c \left[ \frac{m_0}{m_c(E)} g_l^{ '}(a) +\frac{l+1}{a}\Delta m_c(E) 
g_l(a) \right] \, , \quad l=1,2 \dots \, . 
\end{eqnarray}
The conventionally used SBCs for the impenetrable barrier
assumes vanishing of the wave function $\Psi^c$ at the NC
surface,  and thus correspond to $ \gamma_c=0$. In general, however,
the conduction band component of the wave function  does not vanish
at the surface.  Furthermore, it has been shown in
Ref.\onlinecite{bc}  that the BC with $ \gamma_c=0$ is
incorrect in cases where inter-band coupling is important. The use of the GBCs of Eqs. (\ref{bcfso})  gives the equations for the conduction electron energies $E_l^\pm$ of the states with orbital momentum $l$ and total momentum $j_c=l \pm 1/2$ respectively:
\begin{equation}
 E_l^\pm - \frac{E_g}{2} =\frac{\hbar^2 \phi_{l,n}^{\pm 2}}{2m_c(E_l^\pm)a^2} \, ,
  \label{bcl} 
\end{equation}
where $\phi_{l,n}^{\pm}$ is the solution of following equation:
 \begin{equation}
 j_{l}(\phi_{l,n}^\pm) \left[1 - \delta_{cl}^{\pm}\right] =   
\frac{ \gamma_c}{a} \frac{m_0}{m_c(E_l^\pm)}
 \phi_{l,n}^\pm \left[
\frac{l}{2l+1} j_{l-1}(\phi_{l,n}) - \frac{l+1}{2l+1} j_{l+1}(\phi_{l,n})
\right] \,,   
\label{bcphi}
\end{equation}
where $\delta_{cl}^{+} = - l(\gamma_c/a)\Delta m_c(E_l^+)$ for $l=0,1,2 \dots$  and  $j_c=l+1/2$; and $\delta_{cl}^{-} = (l+1)(\gamma_c/a)\Delta m_c(E_l^-)$ for $l=1,2,3 \dots$, and $j_c=l-1/2$.

The $n^{\rm th}$ solution of these equations defines the energy of the
$n^{\rm th}$ electron level with the angular momentum $l$ and the
total angular momentum $j_c=l\pm 1/2$. In the case of the SBCs for the electron component only we can use
$ \gamma_c=0$, and the solution $\phi_{l,n}^\pm$ is the $n$--th zero
$\phi_{l,n}^0$ of the spherical Bessel functions $j_{l}$. The
GBCs with $\gamma_c \ne 0$ takes into account the effect of the surface on
the electron energy levels.

\begin{figure}[p]
\centering
\captionsetup{justification=raggedright,singlelinecheck=false}
\includegraphics[height=0.72\textheight,width=\textwidth,keepaspectratio]{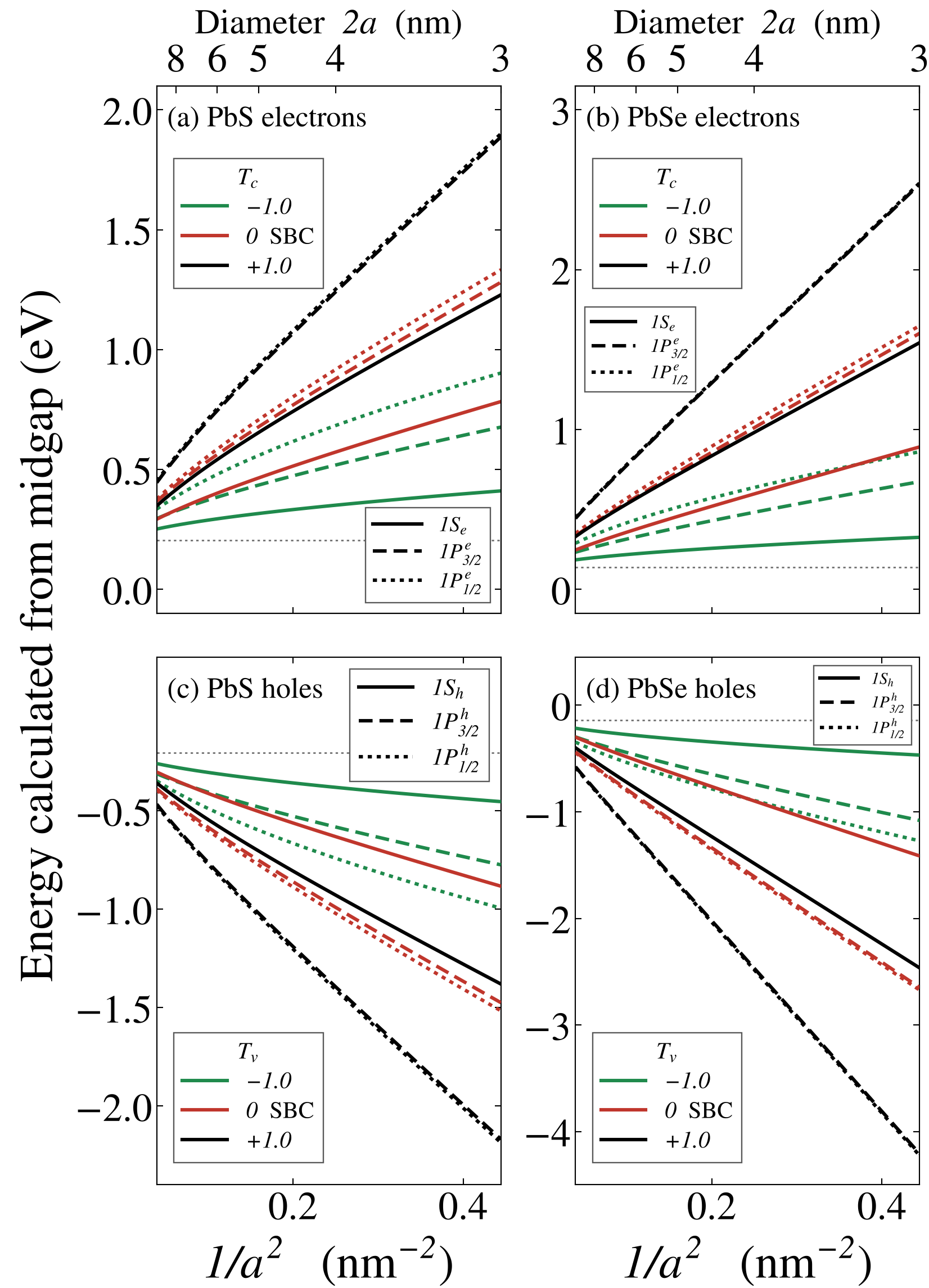}
\caption{Effect of $T_c$ and $T_v$ on the lowest quantum confined  levels in PbS (panel (a) and  (c)) and PbSe  (panel (b) and  (d))  calculated  for 
$\alpha_v = 0$ and  $T_c = 0, \pm 1$ (electrons, upper panels), and for $\alpha_c = 0$ and $T_v = 0, \pm 1$ (holes, lower panels). Panels a(c) and b(d), show the size dependence of $1S$, $1P_{3/2}$ and  $1P_{1/2}$ electron (hole) levels. The surface parameters  $T_c = 0$ and $T_v = 0$ correspond to  the SBC. Colour represents the value of $T_c$ (upper panels) or $T_v$ (lower panels) and line style represents the transition. Green,  red and black lines are calculated for  $T_{c,v} = -1$, 
$T_{c,v} = 0$,  and  $T_{c,v} = +1$, correspondingly.  The confinement electron  and hole levels $1S_{e,h}$,  $1P^{e,h}_{3/2}$  and  $1P^{e,h}_{1/2}$ are shown by solid, 
 dashed and  dotted line  correspondingly.  The horizontal dotted line shows  the bulk
energy gap.}
\label{Fig:2}
\end{figure}

Figure \ref{Fig:2}(a,b) shows the effect of the GBCs on the conduction electron  energy levels  $E^c > E_g/2$  of the ground $1S^c$ ($l=0$) and the first excited $1P^c$ ($l=1$)
 states in bare PbSe NCs calculated with $\alpha_v=0$. The surface induced
spin-orbit interaction splits the last state into two states with
the total angular momentum $j_c=3/2$ ($P_{3/2}^c$ state) and $j_c=1/2$ ($P_{1/2}^c$
state) and the size dependence of this splitting
$\Delta_c=E(1P^c_{3/2})-E(1P^c_{1/2})$ is also seen in Fig. \ref{Fig:2},
 One can see that even a small surface
parameter introduced by the GBCs, $T_c$,  significantly
affects both ground and excited states in small NCs. The
positive (negative) value of the surface parameter $T_c >0$
($T_c<0$) increases (decreases) the energy of all states in
comparison with $ T_c=0$.    For positive values of the surface parameters $T_c$ the GBCs predict the formation of surface (gap) states with energies within the band gap, $-E_g/2 < E < E_g/2$, as was first demonstrated in Ref. \onlinecite{bc}.

 The
value of the spin-orbit splitting, $\Delta_c$, of the states with $E^c > E_g/2$,  has the same sign but larger than those calculated in Ref.(\onlinecite{kang_wise}). However, for
all excited states  with $E^c > E_g/2$ it remains smaller than the averaged quantization energy
$|E_l^c-E_g/2|=|(lE_{l-1/2}+(l+1)E_{l+1/2})/(2l+1)- E_g/2|$ of the level  with the
orbital momentum $l$.  
If $|\delta_{cl}^\pm| \ll 1$, the averaged energies $E_l^c$ and
corresponding wave numbers $\phi_{l,n}$ can be found from the
simplified BCs given by Eq.~(\ref{bcphi}) with $\delta_{cl}^\pm=0$. The  energy corrections  coming from
the small $\delta_{cl}^\pm \ne 0$  can be found perturbatively as shown in Ref. \onlinecite{sopaper}.  Following this method we obtain the following approximate expression for the spin--orbit splitting of the conduction electron energy levels
\begin{eqnarray}
\Delta_c(E_l^c)& = &-\left(2l + 1 \right) \frac{\hbar^2}{2m_0a^2}
\Sigma_{sur}(E_l^c) \, ,
 \label{deltac} \nonumber\\ \Sigma_{sur}(E_l)&= &a^3 \frac{E_p}{E_g/2 +E_l^c} \, 
\frac{{|f_l(a)|}^2l+{|g_l(a)|}^2(l+1)}{2l+1}
\, . \label{sigmasur}
\end{eqnarray}
 Note that although
$\phi_{l,n}^+=\phi_{l,n}^-=\phi_{l,n}$, the normalization constants
$C_l^f \ne C_g^-$ and therefore $|f_l(a)|^2 \ne |g_l(a)|^2$ .  Thus, if $|\Delta_c| \ll |E_l^c-E_g/2|$, the energy levels of the
excited states $E_l^{\pm}$ with $j_c = l \pm 1/2$ are the eigenenergies of the effective Hamiltonian $\hat H_l^c$:
\begin{eqnarray}
\hat H_{l}^c \Omega_{l \pm 1/2,l,m} = E_l^\pm \Omega_{l \pm 1/2,l,m}  \, , \nonumber \\
\hat H_{l}^c = E_l^c\hat U_2 +  \frac{1}{2l+1} \Delta_c (E_l^c)  \left( \mbox{\boldmath $
\hat L \hat  \sigma$} \right) \, .
\label{hlso}
\end{eqnarray}

The dimensionless parameter $\Sigma_{sur}$ is proportional to the  mean value of the square
of the wave function at the surface. It describes the
surface--induced spin--orbit coupling of the conduction electron QCLs
in   NCs.
Thus, the spin--orbit splitting of the electron QCLs  in NCs is caused by the
admixture of the valence band states near the surface and it is significantly enhanced as it is seen in Fig. \ref{Fig:2} (a,b) in the case of the GBCs when the electron component of the wave function does not vanish at the surface.

The effect of the surface  on the size dependence of QCLs of the valence band electrons (holes)  may also be considered in the other analytical limit that is realized  in the case $\alpha_c=0$.  The analysis in Appendix \ref{app:valence} is identical to the one done above for the case $\alpha_v=0$. The summary of this analysis is shown in Fig.~\ref{Fig:2}(c,d), which shows the effect of $T_v$ on the size dependence of QCLs at $T_v=0$.

\section{\label{fullmodel} General boundary condition in the  full  model} 

Let us consider the effect of the GBCs of Eq. (\ref{sphBC}) on  electron and hole energy spectra in the full model with $\alpha_c \alpha_v >0 $. To derive the GBCs for the radial functions $f_l$, $f_{l+1}$, $g_l$ and $g_{l+1}$ on the spherical surface $r=a$ we substitute  the bi-spinor functions of Eq.(\ref{eq:29}) into the GBC of Eq.(\ref{sphBC}) and use the property of the spherical spinors:
\begin{eqnarray}
({\bm \sigma}{\bm \tau}) \Omega_{j,l,m} = (i)^{2j-2l} \Omega_{j,2j-l,m} \, . 
\end{eqnarray}
 We obtain the following for the states with $j=l+1/2$:
\begin{eqnarray}
f_l(a) = \gamma_c \left[ \alpha^+\frac{\partial f_{l}(a)}{\partial r} + \upsilon  f_{l+1}(a) \right] \, , \nonumber \\
f_{l+1}(a) = \gamma_v \left[ \alpha^+\frac{\partial f_{l+1}(a)}{\partial r} + \upsilon f_{l}(a) \right] \, ,
\end{eqnarray}
where $\gamma_c=T_c a^*= T_c/\upsilon$ and $\gamma_v=T_v a^*= T_v/\upsilon$ are two independent surface parameters, and
\begin{eqnarray}
g_l(a) = \gamma_c \left[ \alpha^+\frac{\partial g_{l}(a)}{\partial r} + \upsilon g_{l-1}(a) \right] \, , \nonumber \\
g_{l-1}(a) = \gamma_v \left[ \alpha^+\frac{\partial g_{l-1}(a)}{\partial r} + \upsilon g_{l}(a) \right] \, 
\end{eqnarray}
for the states with $j=l-1/2$. Substituting of the wave functions from Eqs. (\ref{eq:33f},\ref{eq:33A}) in these GBC leads to the systems of the linear equations for the coefficients $C_{l1}^{f,g}$ and $C_{l2}^{f,g}$:
\begin{eqnarray}
C_{l1}^f \left[ j_l(\varphi_1)(1-T_cSl)+j_{l+1}(\varphi_1)T_c(\varphi_1 S-\Lambda(k_1) )\right] + \nonumber \\ 
 C_{l2}^f \left[ j_l(\varphi_2)(1-T_cSl)+j_{l+1}(\varphi_2)T_c(\varphi_2 S-\Lambda(k_2)) \right]  = 0 \, , \nonumber\\
C_{l1}^f \left[ j_l(\varphi_1)T_v (S \varphi_1 \Lambda(k_1)+1)-j_{l+1}(\varphi_1)\Lambda(k_1) (1+T_vS(l+2)) \right] + \nonumber \\ 
C_{l2}^f \left[ j_l(\varphi_2)T_v (S \varphi_2 \Lambda(k_2)+1)-j_{l+1}(\varphi_2)\Lambda (k_2)(1+T_vS(l+2)) \right]= 0 \, , \label{eq:42}
\end{eqnarray}
where $\varphi_1 = k_1a$,  $\varphi_2 = k_2a$,  $S=(\alpha_c + \alpha_v)a^*/a$, and 
\begin{eqnarray}
C_{l1}^g \left[ j_l(\varphi_1)(1+T_cS(l+1))+j_{l-1}(\varphi_1)T_c(\Lambda(k_1)-\varphi_1 S )\right] + \nonumber \\ 
C_{l2}^g \left[ j_l(\varphi_2)(1+T_cS(l+1))+j_{l-1}(\varphi_2)T_c(\Lambda(k_2)-\varphi_2 S )\right]  = 0 \, , \nonumber\\
C_{l1}^g \left[ j_l(\varphi_1)T_v (S \varphi_1 \Lambda(k_1)+1)+
j_{l-1}(\varphi_1)\Lambda(k_1) (1-T_vS(l-1)) \right] + \nonumber \\ 
C_{l2}^g \left[ j_l(\varphi_2)T_v (S \varphi_2 \Lambda(k_2)+1)+
j_{l-1}(\varphi_2)\Lambda(k_2) (1-T_vS(l-1)) \right]=0 \, , \label{eq:48}
\end{eqnarray}
The SBC correspond to $T_c=T_v=0$. With these parameters the Eqs. (\ref{eq:42},\ref{eq:48}) can be solved only if $k_1 \ne k_2$ and thus $\alpha_c \alpha_v \ne 0$. With arbitrarily chosen $T_c$ and $T_v$ (including $T_c=T_v=0$ or $T_cT_v=1$) we obtain the following equations for the energy levels in the case $\alpha_c \alpha_v > 0$ ($k_1 \ne k_2$):
\begin{eqnarray}
&&j_l(\varphi_1)j_l(\varphi_2) T_v S (1-T_cSl)(\Lambda(k_1)\varphi_1-\Lambda(k_2)\varphi_2) + \nonumber\\&&   
j_{l+1}(\varphi_1)j_{l+1}(\varphi_2) T_c S (1+T_vS(l+2))(\Lambda(k_2)\varphi_1-\Lambda(k_1)\varphi_2) + \nonumber \\&&
j_l(\varphi_1)j_{l+1}(\varphi_2) \left[(1-T_cSl)(1+T_vS(l+2))\Lambda(k_2)
+T_cT_v(S\varphi_2-\Lambda(k_2))(\Lambda(k_1)S\varphi_1  +1)\right] + \nonumber \\&&
j_{l+1}(\varphi_1)j_{l}(\varphi_2) \left[(T_cSl-1)(1+T_vS(l+2))\Lambda(k_1)
+T_cT_v(S\varphi_2\Lambda(k_2)+1)(\Lambda(k_1)-S\varphi_1)\right] = 0 \, , \nonumber\\
\end{eqnarray}
for the states with total momentum $j=l+1/2$ and $l=0,1,\dots$; and 
\begin{eqnarray}
&&j_l(\varphi_1)j_l(\varphi_2) T_v S (1+T_cS(l+1))(\Lambda(k_2)\varphi_2-\Lambda(k_1)\varphi_1) + \nonumber\\&&   
j_{l-1}(\varphi_1)j_{l-1}(\varphi_2) T_c S (1-T_vS(l-1))(\Lambda(k_1)\varphi_2-\Lambda(k_2)\varphi_1) + \nonumber \\&&
j_l(\varphi_1)j_{l-1}(\varphi_2) \left[(1+T_cS(l+1))(1-T_vS(l-1))\Lambda(k_2)
+T_cT_v(S\varphi_2-\Lambda(k_2))(\Lambda(k_1)S\varphi_1  +1)\right] + \nonumber \\&&
j_{l-1}(\varphi_1)j_{l}(\varphi_2) \left[(T_cS(l+1)+1)(T_vS(l-1)-1)\Lambda(k_1)
+T_cT_v(S\varphi_2\Lambda(k_2)+1)(\Lambda(k_1)-S\varphi_1)\right] = 0 \, , \nonumber\\
\end{eqnarray}
for the states with total momentum $j=l-1/2$ and $l=1,2,\dots$.

\begin{figure}[H]
\centering
\includegraphics[width=7.5cm]{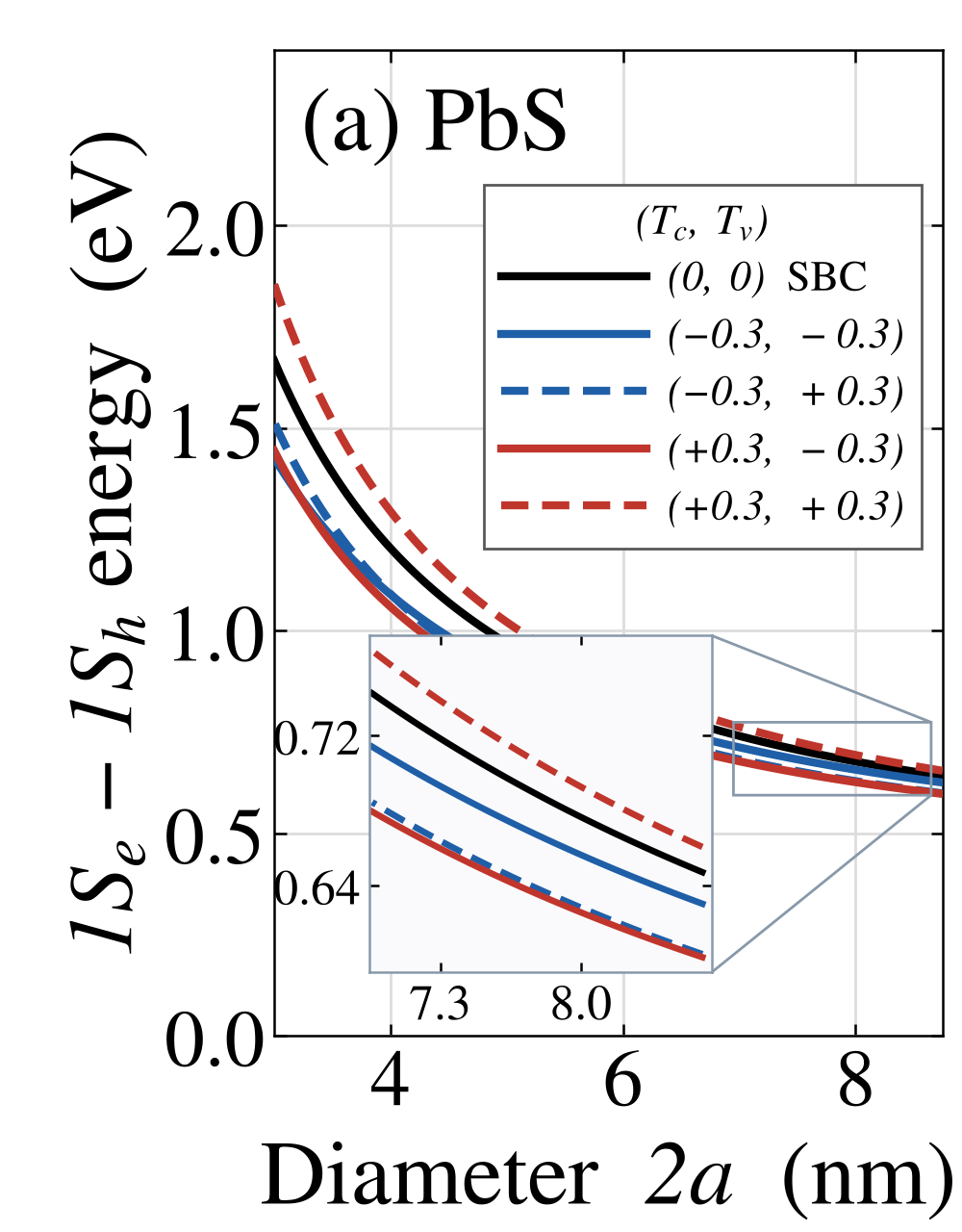}
\qquad
\includegraphics[width=7.5cm]{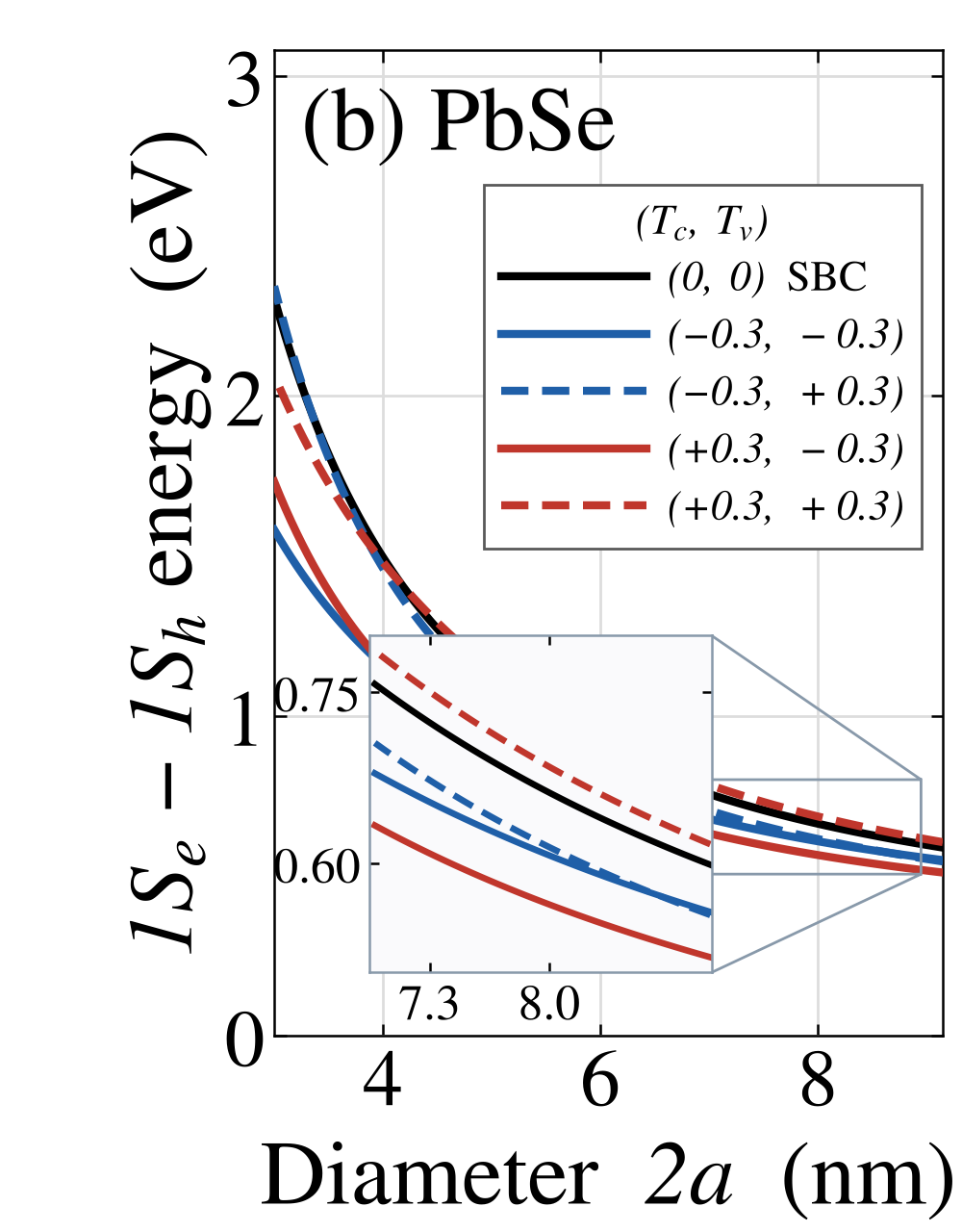}
\caption{Combined effect of the two independent surface parameters on the
size dependence of the lowest $1S_e-1S_h$ optical transition, calculated
in PbS (a) and PbSe (b) NCs  for the four sign combinations of
$T_c = \pm 0.3$ and $T_v = \pm 0.3$, together with the SBC $T_c = T_v = 0$  for SBC (the black line).  Blue  and red line  correspond to  $T_c = -0.3$ and  $T_c = +0.3$. ; solid lines have $T_v = -0.3$ and dashed lines have
$T_v = +0.3$. The four combinations are therefore blue solid
$(-0.3,-0.3)$, blue dashed $(-0.3,+0.3)$, red solid $(+0.3,-0.3)$ and
red dashed $(+0.3,+0.3)$. The inserted window magnifies the large-size end, where
the five curves converge.}
\label{level_0}
\end{figure}

 Figure \ref{level_0} shows the combined effect of the two surface parameters $T_c$ and $T_v$ on the size dependence of the lowest $1S_e-1S_h$ optical transition in PbS and PbSe NCs. The analysis presented above describes the QCLs in the conduction ($E>E_g/2$) and valence ($E<-E_g/2$) bands and show  the effect of the boundary
conditions.  
The GBCs predict the possibility of surface (gap) states forming within the band gap, $-E_g/2 < E < E_g/2$, for positive values of the surface parameters $T_c$ and $T_v$. This effect of the GBCs was first demonstrated in Ref.  \onlinecite{bc}. These gap states may influence the photoluminescence properties of NCs and will be investigated in future work. 
\section{\label{absorption} Absorption spectra of PbS, PbSe  and Cl passivated PbS  nanocrystals } 
In this section we  demonstrate that  GBCs allow us to quantitatively describe the size dependence of optical transition energies measured in PbS  and PbSe NCs.  We also describe the size dependent absorption spectra of Cl passivated PbS NCs, which show a significant blue shift of the optical transition energies.\cite{BrittmanPCL2019} Our GBCs allow us to model this shift and provide some insight into the effect of Cl on the photoluminescence quantum yield through the modification of the electron and hole wave functions.

\begin{figure}[H]
\centering
\includegraphics[width= 1.0\textwidth]{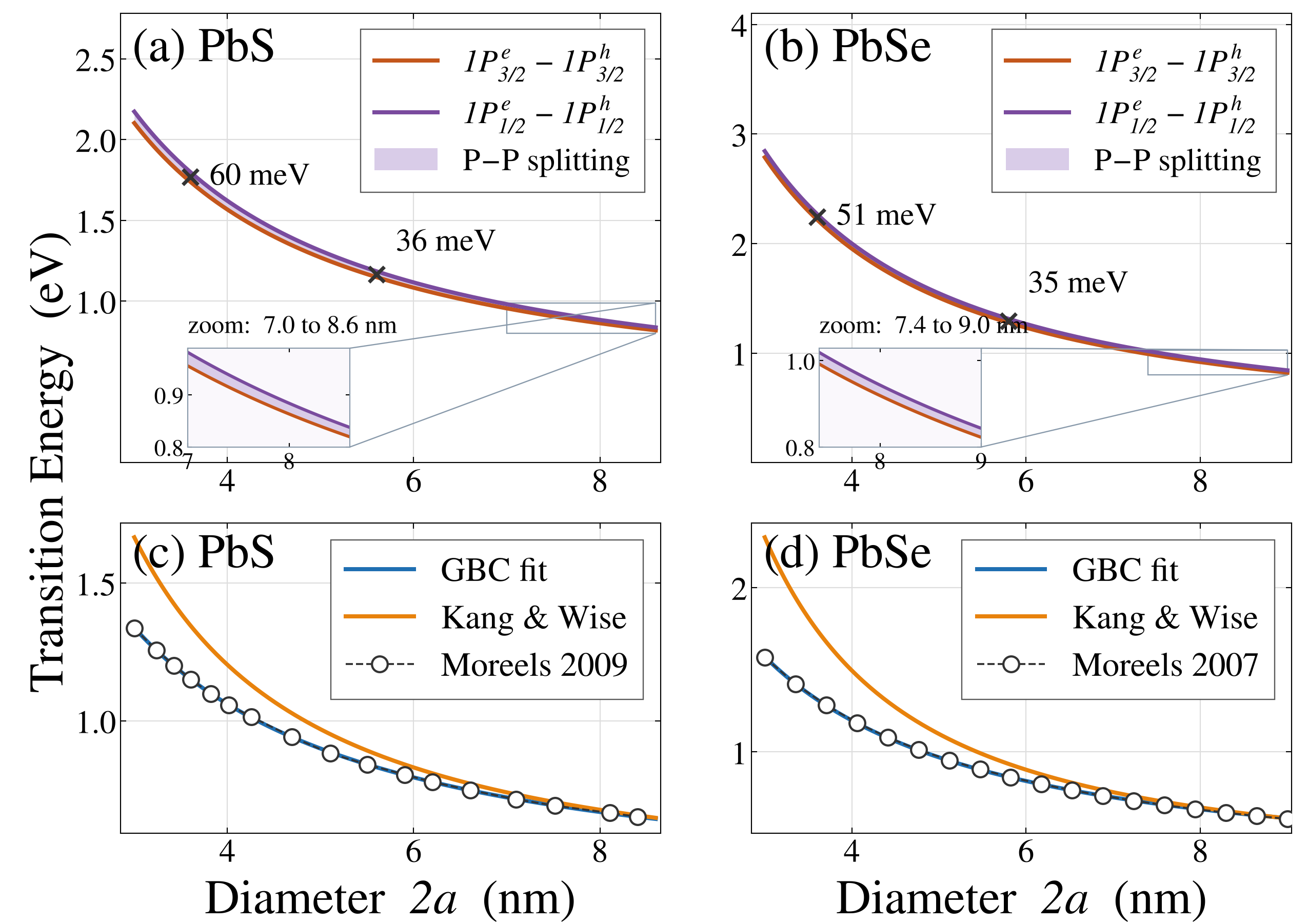}
\caption{The size dependence of lowest optical transitions in bare   PbS and PbSe NCs.       On panel (c)  the blue line that goes through Moreels data \cite{Moreels}  measured in PbS NC  is calculated using GBCs with  $T_c=-0.825$ and $T_v=-0.525$. The fit RMS=1.6\,meV.  On the panel (d)  the blue line that goes through  Moreels data \cite{Moreels2007}  measured in PbSe NCs is calculated using GBCs with  $T_c=-0.455$ and $T_v=-0.410$. The fit RMS=2.5\,meV.  The yellow lines  show the size dependence of the lowest  $1S_e-1S_h$  optical transition calculated with SBC ($T_c=T_v=0$) and correspond to the Kang--Wise model\cite{kang_wise}.  Using the GBCs with $T_c$ and $T_v$ extracted  from the size dependence of the ground  $1S_e-1S_h$ optical transition we calculated size dependence of  $1P^e_{3/2}-1P^h_{3/2}$ and $1P^e_{1/2}-1P^h_{1/2}$  transitions in bare PbS (a)and PbSe (b) NCs. Inserted windows magnify the spin-orbit splitting between them. }
\label{Fig.4}
\end{figure}

We start with a description of the ground state optical transition in bare PbS NCs, which is described by the $universal$ Moreels curve  shown in Fig.\ref{Fig.4} (c) \cite{Moreels,Cademartiri}  and  bare PbSe  NCs shown in Fig.~\ref{Fig.4}(d)   \cite{Moreels2007}. The experimental results are compared with calculations conducted in the Kang--Wise model \cite{kang_wise} and in the GBC model of this paper.

Figure \ref{Fig.4} shows clearly that the Kang--Wise model corresponding to SBCs with $T_c=T_v=0$  significantly overestimated the optical transition energies in both PbS and PbSe NCs. At the same time, the GBC model with negative $T_c$ and $T_v$ shows excellent agreement with the size dependence of the ground state $1S_h-1S_e$ transitions in PbS NCs using surface parameters $T_c=-0.825$ ($\gamma_c=-0.10$\,nm) and $T_v=-0.525$ ($\gamma_v=-0.08$\,nm)  and in PbSe NCs using  $T_c=-0.455$  ($\gamma_c=-0.051$\,nm) and $T_v=-0.410$ ($\gamma_v=-0.049$\,nm). The NC sizes reported in Refs. \cite{Moreels,Cademartiri,Moreels2007} were determined using transmission electron microscopy (TEM). Small-angle X-ray scattering (SAXS) measurements yield slightly different NC sizes, resulting in a correspondingly different size dependence of the optical transition energies. Recently, the size dependence of the lowest-energy optical transitions in PbS and PbSe NCs, with NC sizes determined by SAXS, was reported in Ref. \onlinecite{NanoLetter22}. We find that these experimental dependences can be quantitatively described within the GBC model using the following phenomenological surface parameters: PbS NCs $T_c=-0.147$ ($\gamma_c=-0.018$\,nm) and $T_v=-0.184$
($\gamma_v=-0.023$\,nm), and in PbSe NCs $T_c=-0.653$
($\gamma_c=-0.080$\,nm) and $T_v=-0.626$ ($\gamma_v=-0.076$\,nm).

\begin{figure}[H]
\centering
\includegraphics[width= 0.9\textwidth]{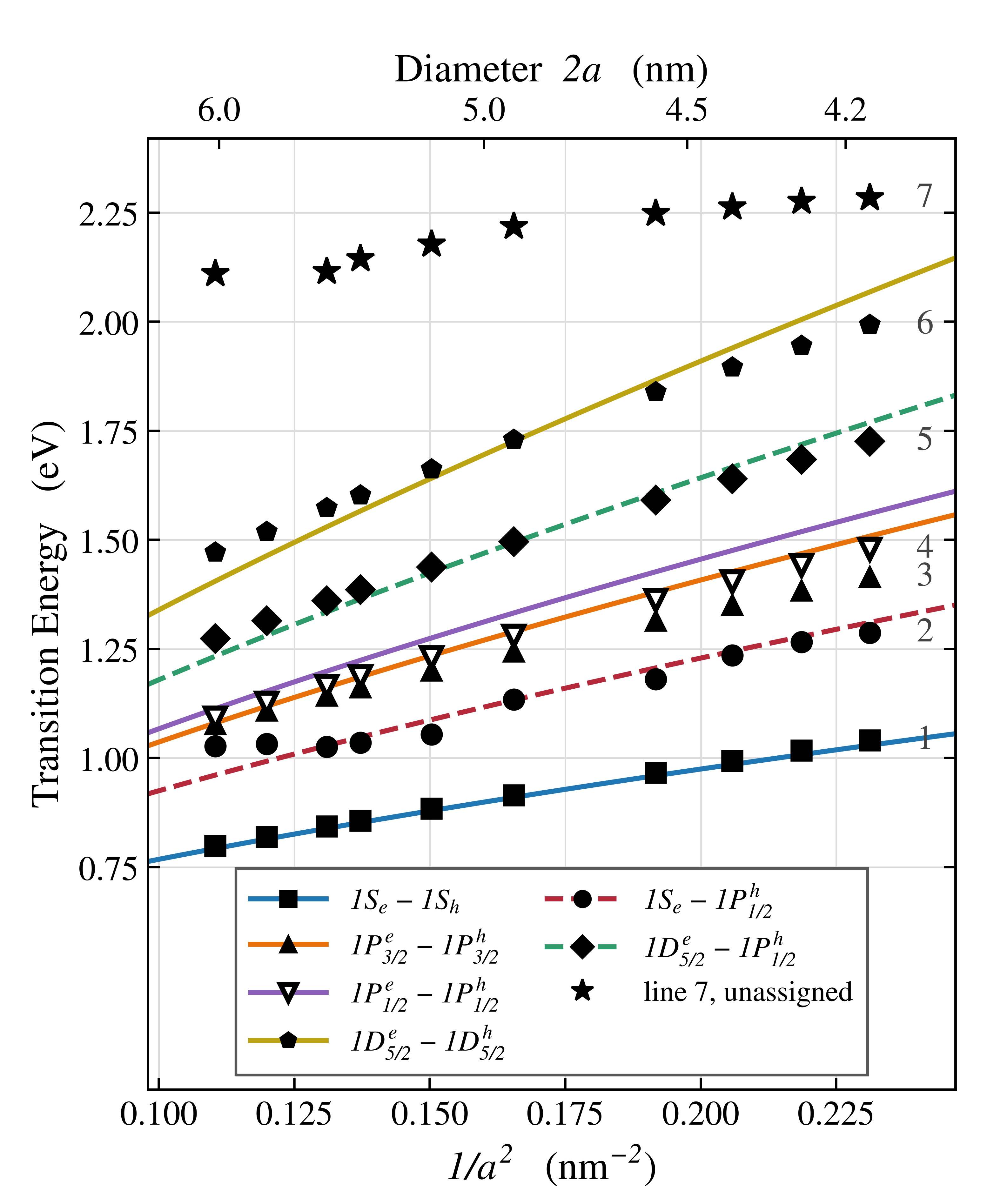}
\caption{The size dependence of  the lowest optical transitions reported in Ref. \cite{pbs}. The experimental   data  that presented in Ref. \cite{pbs} as $E_n$  lines where  $n=1,2,3.4,5,~{\rm and} ~6$  can described as four $1S_h-1S_e$,  $1P_{3/2}^h-1P_{3/2}^e$, $1P_{1/2}^h-1P_{1/2}^e$  and $1D_{5/2}^h-1D_{5/2}^e$  optically allowed and two $1S_e-1P_{1/2}^h$ and  $1D_{5/2}^e-1P_{1/2}^e$  optically forbidden transitions using GBC with $T_c=-0.825$ and $T_v=-0.525$. These optical transition curves were calculated using parameter extracted  from the size dependence of the ground state transition shown  in Fig. \ref{Fig.4} (c).}
\label{Fig.5}
\end{figure}

Using  the GBCs with PbS surface parameters $T_c=-0.825$ ($\gamma_c=-0.10$\,nm) and $T_v=-0.525$ ($\gamma_v=-0.08$\,nm)  extracted from the fitting  of the lowest $1S_h-1S_e$ optical transitions  measured in Ref. \cite{Moreels}, we were able to identify and quantitatively describe the size dependence of the excited optical transition reported in Ref. \cite{pbs}.   
Figure \ref{Fig.5} shows the size dependence of the optical transitions measured in Ref.  \cite{pbs}  and our description of these transitions calculated  using GBCs.  The first six transitions are very well described as four $1S_h-1S_e$,  $1P_{3/2}^h-1P_{3/2}^e$, $1P_{1/2}^h-1P_{1/2}^e$  and $1D_{5/2}^h-1D_{5/2}^e$  optically allowed and two $1S_e-1P_{1/2}^h$ and  $1D_{5/2}^e-1P_{1/2}^e$  optically forbidden   transitions.  The upper 7th transition obviously does not converge to the bulk band edge with increasing NC size and belongs to the series of optical transitions originating from the $\Sigma$ critical point of the Brillouin zone. \cite{pbs,pbse,PbSNanorod}

The experimental lines in Ref.  \cite{pbs} are quite broad, which could be caused by the deviation of the NCs structure from the  spherical symmetry. Also the number of the optically allowed  and optically forbidden transitions  increases   due to an increase of total angular momenta $j$. In the case of the $D$ orbital state $j$ can be 5/2, 3/2, and 1/2, our calculations of the optical transitions are conducted in spherical approximations and  TEM images of NCs shown in Ref.  \cite{pbs}. These images clearly show asymmetry of the NCs shape. The shape anisotropy oriented along some crystallographic direction, combined with anisotropy  of effective masses in the PbS valley, \cite{Poddubny2017} can potentially explain why the blue shift of the $1P_{3/2}^h-1P_{3/2}^e$, $1P_{1/2}^h-1P_{1/2}^e$, and $1D_{5/2}^h-1D_{5/2}^e$ optical transitions slows down with decreasing NC size compared to the spherical approximation. The nonspherical shape of NCs can also break  the spatial inversion symmetry and  explain the observation of the forbidden transition between the states with the opposite parity of the envelope functions.   Uncertainty  about the NC shape  and dispersion   prevents us from doing a global fit of the experimental data.

The Cl passivated PbS NCs show a significant blue shift of the ground-state optical  $1S_e1S_h$ transition. They also show an increase of the NC lattice constant with decrease of NC radius, \cite{BrittmanPCL2019} which can be induced by surface Cl stretching. Within the phenomenological continuous media model  this increase in  spherical NCs  can be described as  $\beta/a$, where $\beta$ is the surface tension coefficient. The increase of the lattice constant, usually associated with rising temperature, is known to increase the bulk energy gap in PbS and PbSe.  As a result, the surface induced NC tension modifies the NC energy gap according to $\alpha/a$, where $a$ is the NC radius.

The authors of Ref.  \cite{BrittmanPCL2019} also observed a second optical transition above the ground $1S_e1S_h$ transition.  The origin of the second transition is still under discussion.  It has been suggested that, in addition to the optically allowed $1P_{3/2}^e1P_{3/2}^h$ and $1P_{1/2}^e1P_{1/2}^h$ transitions, the parity forbidden $1S_e1P_e$ and $1S_h1P_h$ transitions can also be observed in PbS and PbSe NCs.

\begin{figure}[H]
\centering
\includegraphics[width=0.8\textwidth]{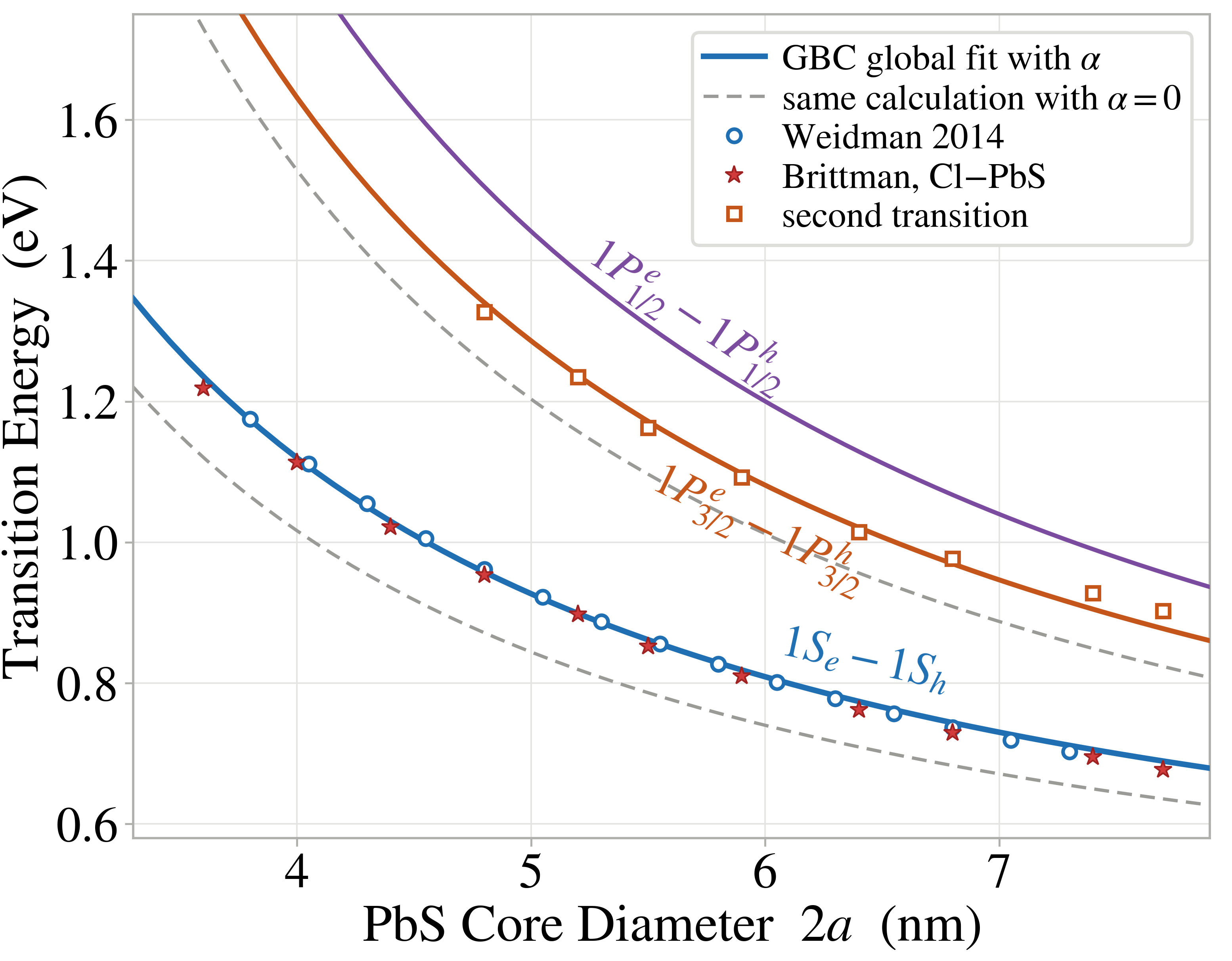}%
\caption{Transition energy measured as a function of the core diameter of PbS NCs. Experimental size dependence (Refs. \cite{Weidman2014,BrittmanPCL2019}) of the lowest optical transitions $1S_e-1S_h$ and $1P_{3/2}^e-1P_{3/2}^h$  and its global fit using GBC. The best fit was reached for $T_c=0.12$ and $T_v= -1.01$ and surface tension parameter  $\alpha=-3.90$eV$\cdot a_0$. Figure also shows the size dependence of the  $1P_{1/2}^e-1P_{1/2}^h$ calculated theoretically. }%
\label{fig.6}
 \end{figure}

 The universal fit (Fig. \ref{fig.6}) of the first ($1S_e-1S_h$) and second ($1P_{3/2}^e-1P_{3/2}^h$) optical transitions using  GBCs with $T_c=0.12$ ($\gamma_c=0.014$\,nm) and $T_v=-1.01$  ($\gamma_v=-0.125$\,nm) and surface tension parameter  $\alpha=-3.90$eV$\cdot a_0$, where $a_0$ is the Bohr radius, provides their excellent description  with RMS = 8.4\,meV.  The assumption that the second observed transition is connected with $1S_e1P_h$ or $1S_h1P_e$ parity-forbidden transitions increases the RMS to 41.7 or 54.1\,meV, roughly three times the 14.7\,meV obtained for the $1P_e1P_h$ assignment. The calculations also show that the surface effect introduced by GBC significantly increases $1P_{3/2}^e1P_{3/2}^h$ and $1P_{1/2}^e1P_{1/2}$ splitting.


\section{\label{sum}Discussion and Conclusion}

Using GBCs, we were able to quantitatively describe the size dependence of the lowest optical transitions in bare PbS and PbSe NCs. The two phenomenological parameters of our theory, $T_c$ and $T_v$, used to describe the experimental data were found to be negative. Using the earlier approach to GBCs developed in Ref.~\onlinecite{bc}, we interpret these results as indicating that the electronic NC radius is larger than the ionic NC radius, $a$, measured by TEM. In contrast, in Cl-passivated PbS NCs, where $T_c$ was found to be positive, the electronic NC radius should be smaller than the ionic NC radius, $a$. This reduction of the effective electronic NC radius naturally increases the optical transition energies.

 As was shown in Ref.~\cite{gupta02}, surface-induced spin--orbit coupling leads to an additional contribution to the electron $g$-factor in CdSe NCs. Similar spin--orbit terms exist in PbS and PbSe NCs, as shown in Section IV. These terms should give rise to surface contributions to the electron and hole $g$-factors in PbS and PbSe NCs, whose magnitudes are proportional to the squares of the surface parameters $T_c$ and $T_v$, respectively. The difference between the surface parameters in bare and Cl-passivated PbS NCs indicates that the $g$-factors and their size dependence in these NCs should also be different.

Finally, the reduction of the effective electron and hole radii suggests additional confinement experienced by the carriers in Cl-passivated PbS NCs. This additional confinement may potentially contribute to the high photoluminescence QY observed in Cl-passivated PbS NCs \cite{BrittmanPCL2019}, because it suppresses the escape of photoexcited carriers into the surrounding matrix. As a result, the NCs remain neutral under optical excitation, thereby suppressing nonradiative Auger recombination that would otherwise quench the radiative emission.

{\em In summary}, we have shown that the semiconductor surface has an important influence on the electron energy structure in bare spherical lead-halide NCs. The surface effect has been modeled through the choice of the boundary-condition parameter that accounts for the nonzero value of the envelope function at the NC surface. We have shown that the surface induces additional nonparabolicity of the quantum-size energy levels,  and spin--orbit splitting of the electron and hole quantum-confined  levels. These effects are particularly significant in small NCs. Analysis of the experimental data for PbS NCs allows us to determine the surface parameters of the GBC for bare PbS and PbSe NCs and for Cl-passivated PbS NCs.

\begin{acknowledgments}
The authors thank M. W. Swift for his help with recovery of the old Mathematica code,  A.M. Minarskii for useful discussions and S. F.  Brittman for the  critical discussion of experimental data published in her paper. The work of A.V.R. was  supported by the  Russian Science Foundation under project No. 23-12-00300-$\Pi$.  Al. L. E. thanks  financial support of the U.S.
 Office of Naval Research (ONR). O. L. thanks the hospitality of Naval  Research Laboratory, where he spent the summer of 2026 as a volunteer.  
\end{acknowledgments}


\appendix
\section{Electron and Hole Confinement  Levels in the Approximate Model  with $\alpha_c = 0$}
\label{app:valence}

Let us consider surface effects in the approximate model $\alpha_c = 0$. We obtain equations for the valence electron energies $E_\lambda^\pm$ of the states with orbital momentum $\lambda$ and total momentum $j_v=\lambda \pm 1/2$, respectively. The valence band electron levels in this approximation can be written as:
\begin{equation}
 E_\lambda^\pm = - \frac{E_g}{2} - \frac{\hbar^2 (\phi_{\lambda,n}^{\pm})^2}{2m_v(E_\lambda^\pm)a^2} \, ,
 \label{bvl}
\end{equation} 
where $\phi_{\lambda,n}^{\pm}$ are the solutions of the following equation:
\begin{equation}
 j_{\lambda}(\phi_{\lambda,n}^\pm) \left[1 - \delta_{v\lambda}^{\pm}\right] =   
\frac{\gamma_v}{a} \frac{m_0}{m_v(E_\lambda^\pm)}
 \phi_{\lambda,n}^\pm \left[
\frac{\lambda}{2\lambda+1} j_{\lambda-1}(\phi_{\lambda,n}^\pm) - \frac{\lambda+1}{2\lambda+1} j_{\lambda+1}(\phi_{\lambda,n}^\pm)
\right] .  
\label{bvphi}
\end{equation}
Here
\begin{equation}
\delta_{v\lambda}^{+} =  \lambda\, \frac{\gamma_v}{a}\Delta m_v(E_\lambda^+) \, , \quad \lambda=0,1,2 \dots \, ,
\end{equation}
for the states with $j_v=\lambda+1/2$, and
\begin{equation}
\delta_{v\lambda}^{-} = -(\lambda+1)\,\frac{\gamma_v}{a}\Delta m_v(E_\lambda^-) \, , \quad \lambda=1,2,3 \dots \, ,
\end{equation}
for the states with $j_v=\lambda-1/2$.
The $n^{\rm th}$ solution of these equations defines the energy of the $n^{\rm th}$ electron level with angular momentum $\lambda$ and total angular momentum $j_v=\lambda \pm 1/2$. In the case of the SBCs, $\gamma_v=0$, and the solution $\phi_{\lambda,n}^\pm$ is the $n^{\rm th}$ zero $\phi_{\lambda,n}^0$ of the spherical Bessel function $j_{\lambda}$. The GBCs with $\gamma_v \ne 0$ take into account the effect of the surface on the valence electron energy levels.

The effect of the GBCs on the valence band hole energy levels $E^v < -E_g/2$ of the ground $1S^v$ ($\lambda=0$) and first excited $1P^v$ ($\lambda=1$) states in bare PbS and PbSe NCs calculated with $\alpha_c=0$ is shown in Fig.~\ref{Fig:2}(c,d). The surface-induced spin--orbit interaction splits the excited state into two states with total angular momentum $j_v=3/2$ ($P_{3/2}^v$ state) and $j_v=1/2$ ($P_{1/2}^v$ state); the size dependence of this splitting $\Delta_v = E(1P^v_{3/2}) - E(1P^v_{1/2})$ is shown in Fig.~\ref{Fig:2}(c,d). Even a small surface parameter introduced by the GBCs significantly affects both ground and excited states in small NCs. A positive (negative) value of the surface parameter $\gamma_v > 0$ ($\gamma_v < 0$) decreases (increases) the energy of all states in comparison with $\gamma_v=0$. For positive values of $\gamma_v$, ground-state energy levels within the forbidden gap $-E_g/2 < E^v < E_g/2$ may appear. In the model with symmetric bands $\alpha_c = \alpha_v = 0$ and symmetric BCs $\gamma_c = \gamma_v$, the calculated energy levels of the conduction and valence band electrons are symmetric with respect to the middle of the band gap.


If $|\delta_{v\lambda}^\pm| \ll 1$, the averaged energies $E_\lambda^v = |(\lambda E_{\lambda-1/2} + (\lambda+1)E_{\lambda+1/2})/(2\lambda+1) + E_g/2|$ of the level with orbital momentum $\lambda$, and the corresponding wave numbers $\phi_{\lambda,n}$, can be found from the simplified BCs given by Eq.~(\ref{bvphi}) with $\delta_{v\lambda}^\pm = 0$. Following the perturbation approach of Ref.~\onlinecite{sopaper}, we obtain the following approximate expression for the spin--orbit splitting of the valence electron energy levels with $E^v < -E_g/2$:
\begin{eqnarray}
\Delta_v(E_\lambda^v) &=& \left(2\lambda + 1 \right) \frac{\hbar^2}{2m_0a^2} \Sigma_{sur}^v(E_\lambda^v) \, , \label{deltav} \\
\Sigma_{sur}^v(E_\lambda^v) &=& a^3 \frac{E_p}{E_g/2 - E_\lambda^v} \frac{{|g_\lambda(a)|}^2\lambda + {|f_\lambda(a)|}^2(\lambda+1)}{2\lambda+1} \, . \label{sigmasur_v}
\end{eqnarray}
Thus, if $|\Delta_v| \ll |E_g/2 - E_\lambda^v|$, the energy levels of the excited states $E_\lambda^{\pm}$ with $j_v = \lambda \pm 1/2$ are the eigenvalues of the effective Hamiltonian $\hat H_\lambda^v$:
\begin{eqnarray}
\hat H_{\lambda}^v \Omega_{\lambda \pm 1/2,\lambda,m} &=& E_\lambda^\pm \Omega_{\lambda \pm 1/2,\lambda,m} \, , \nonumber \\
\hat H_{\lambda}^v &=& E_\lambda^v \hat U_2 + \frac{1}{2\lambda+1} \Delta_v (E_\lambda^v) \left( \bm{\hat L} \cdot \bm{\hat \sigma} \right) .
\label{hlso_v}
\end{eqnarray}
The spin--orbit splitting of the valence electron levels is caused by the admixture of conduction band states near the surface.

\end{document}